\documentclass[11pt]{article}
\usepackage{a4wide}
\usepackage{mathrsfs}
\usepackage{latexsym,bm}
\usepackage{graphicx}
\usepackage{indentfirst}
\usepackage{slashed}
\usepackage{amsmath}
\usepackage{amssymb}
\usepackage[usenames,dvipsnames]{color}
\usepackage{hyperref}
\usepackage{epsfig}
\usepackage[titletoc]{appendix}
\usepackage{multirow}%
\usepackage{rotating}
\usepackage[square,numbers,sort&compress]{natbib}
\usepackage{placeins}   
\usepackage{ulem} 
\usepackage[top=2.9cm,bottom=2.5cm,left=2.8cm,right=3cm]{geometry}
\usepackage{tabularx}

\renewcommand{\arraystretch}{1.0}

\newcommand{\email}[1]{\footnote{{\em } \texttt{#1}}}

\newcommand{\bma}{\left(\begin{matrix}}
\newcommand{\ema}{\end{matrix}\right)}

\newcommand{\mL}{\mathcal{L}}
\newcommand{\mM}{\mathcal{M}}

\newcommand{\rxt}{{\rm R}{\chi}{\rm T}}

\begin{document}
\title{\Large \bf Unified study of two-meson and axion-meson production from semileptonic tau decays within resonance chiral framework}
\author{\small Jin Hao$^a$,\, Chun-Gui Duan$^a$\email{duancg@hebtu.edu.cn},\,  Zhi-Hui Guo$^{a}$\email{zhguo@hebtu.edu.cn} \\[0.5em]
{ \small\it ${}^a$ Department of Physics and Hebei Key Laboratory of Photophysics Research and Application, } \\ 
{\small\it Hebei Normal University,  Shijiazhuang 050024, China}
}
\date{}

\maketitle
\begin{abstract}
We carry out the joint study of the semileptonic tau decays into the two-meson and axion-meson channels, viz. $\tau^-\to (P_1P_2)^-\nu_\tau$ and  $\tau^-\to \pi^-(K^-) a\nu_\tau$ within the framework of resonance chiral theory by including the model-independent axion-gluon-gluon interaction. By utilizing the $\pi^0$-$\eta$-$\eta'$-axion mixing matrix elements from recent studies, we calculate the pertinent two-pseudoscalar boson form factors.  To simultaneously fit the experimental spectra measured in the Cabibbo allowed $\tau^-\to \pi^-\pi^0\nu_\tau$ process and also the Cabibbo suppressed $\tau^-\to(K_S\pi^-, K^-\eta)\nu_\tau$ ones, we determine all the relevant hadron resonance parameters. Then we give predictions to the spectra and branching ratios for various channels, such as $\tau^-\to(\pi^-\eta,\pi^-\eta',K^-\eta',\pi^-a,K^-a)\nu_\tau$. We also calculate the forward-backward asymmetries for all the aforementioned channels. The interplay between the scalar and vector form factors for different observables is analyzed in detail. Our theoretical predictions supply useful guidance to the future tau experiments, such as those at Belle-II, Super Tau-Charm Facility and Tera-Z factory of Circular Electron-Positron Collider.  
\end{abstract}

\section{Introduction}

The longstanding strong CP problem in QCD can be elegantly explained by the Peccei-Quinn (PQ) mechanism~\cite{Peccei:1977hh,Peccei:1977ur}, which predicts an extremely luring particle, viz. the axion~\cite{Weinberg:1977ma,Wilczek:1977pj}. The quest for the latter hypothesized particle has been intensively carried out in a wide scope of physics, including quantum precision measurement, optical cavity, astrophysical observations, rare meson decays and so on~\cite{Kim:2008hd,Graham:2015ouw,GrillidiCortona:2015jxo,Irastorza:2018dyq,Sikivie:2020zpn,Choi:2020rgn,DiLuzio:2020wdo,Cong:2024qly,Jiang:2024boi,MartinCamalich:2025srw}. 

The featured axion interaction is given by the anomalous operator $aG\tilde{G}/f_a$, where $G$ and $\tilde{G}$ represent the gluon field strength tensor and its dual form in order, and $f_a$ is the axion decay constant. Since the $aG\tilde{G}/f_a$ term is responsible for resolving the strong CP problem and is a universal prediction of various ultraviolet axion models, it is usually deemed as the model-independent axion interaction. The anomalous axion-gluon-gluon operator inevitably induces the axion-hadron interactions, which can be probed in axion-baryon scattering~\cite{Carenza:2020cis,Vonk:2021sit,Li:2023thv,Cavan-Piton:2024ayu,Cao:2024cym,Guo:2025icf,Cavan-Piton:2025nsj}, axion-meson reactions~\cite{DiLuzio:2021vjd,DiLuzio:2022gsc,Notari:2022ffe,Wang:2023xny}, light-flavor meson decays~\cite{Bauer:2021wjo,DiLuzio:2022tbb,Alves:2020xhf,Alves:2024dpa,Wang:2024tre,Gao:2022xqz,Gao:2024vkw}, etc. 

It is reminded that tau, as the only lepton that has large enough mass for the hadronic decays, offers a valuable opportunity to study the hadron interactions. Large amounts of tau events are being collected at the ongoing Belle-II experiment~\cite{Belle-II:2018jsg} and could be also abundantly produced in the future experiments, such as Super Tau-Charm Facility (STCF)~\cite{Achasov:2023gey,Cheng:2025kpp,Sang:2020ksa,Bevan:2015nra} and Tera-Z factory at Circular Electron Positron Collider (CEPC)~\cite{CEPCStudyGroup:2018ghi,Ai:2024nmn}. High sensitivity of the rare tau decay channels can be reached at such future facilities. 
Therefore it is interesting to explore the axion-meson interaction in the semileptonic tau decays. Such kind of theoretical study is still rare, and we carry out a pioneer investigation in this work. 

To be specific, we focus on the $\tau^-\to\pi^- a \nu_\tau$ and $\tau^-\to K^- a \nu_\tau$ processes. It is quite plausible that hadronic resonances will play crucial roles in such decays. This requires us to estimate not only axion-light pseudoscalar meson couplings but also the axion-resonance ones. To account for the latter types of couplings, the framework of resonance chiral theory ($\rxt$)~\cite{Ecker:1988te}, together with the $a$-$\pi^0$-$\eta$-$\eta'$ mixing formulas in Refs.~\cite{Gao:2022xqz,Gao:2024vkw}, will be employed in our study. 
Comparing with the general setups of the axion weak chiral Lagrangians in Ref.~\cite{Bauer:2021wjo}, we will stick to the model-independent axion interaction, i.e., the $aG\tilde{G}$ term. The primary aim of the present work is to systematically include the hadronic resonance contributions in the $\tau^-\to\pi^-(K^-) a \nu_\tau$ decays and to investigate to what extent the axion production rates from the tau decays can be enhanced after including the hadronic resonances. To make a self-consistent study, we also update the calculation of the $\tau^-\to (P_1 P_2)^-\nu_\tau$ processes in $\rxt$ both for the Cabibbo allowed and suppressed situations, instead of just focusing upon one specific channel. A joint fit of different kinds of experimental data, including the invariant mass distributions of the $\pi^-\pi^0$, $K_S\pi^-$ and $K^-\eta$ systems from the $\tau^-\to\pi^-\pi^0\nu_\tau$, $\tau^- \to K_S\pi^-\nu_\tau$ and $\tau^-\to K^-\eta\nu_\tau$ decays, respectively, will be performed to determine the unknown resonance couplings. The updated resonance parameters will be further exploited to predict the spectra and branching ratios of the channels that are yet to be measured, such as $\tau^- \to K^-\eta'\nu_\tau$, $\tau^-\to \pi^-\eta(\eta')\nu_\tau$ and $\tau^-\to\pi^-(K^-) a \nu_\tau$. Meanwhile, we also give predictions to the forward-backward asymmetries arising from all the relevant channels, including $\tau^-\to K_S\pi^-\nu_\tau$, $\tau^- \to K^-\eta(\eta')\nu_\tau$, $\tau^-\to \pi^-\eta(\eta')\nu_\tau$ and $\tau^-\to\pi^-(K^-) a \nu_\tau$. 

This article is structured as follows. The general formulas for the theoretical description of the two-boson semileptonic tau decays are given in Sec.~\ref{sec.tauamp}, in order to set up the notations. We elaborate calculations of the vector and scalar form factors within the framework of resonance chiral theory in Sec.~\ref{sec.calcff}. Next, we discuss the combined fit to the experimental data from the $\tau^-\to(\pi^-\pi^0,K_S\pi^-,K^-\eta)\nu_\tau$ decays in Sec.~\ref{sec.fit}. The predictions to the spectra and branching ratios of the $\tau^-\to(\pi^-\eta,\pi^-\eta',K^-\eta',\pi^-a,K^-a)\nu_\tau$ processes are presented in Sec.~\ref{sec.prediction}, where the forward-backward asymmetric distributions are also calculated for all the pertinent channels. We give the summary and conclusions in Sec.~\ref{sec.sum}.

\section{Description of the tau decay amplitudes}\label{sec.tauamp}

The Standard Model (SM) structure for the charged currents (CC) will be employed to calculate the semileptonic tau decays, which can be cast in the conventional form of four-fermion operator as 
\begin{equation}
\mL^{\rm SM}_{\rm CC}= -\frac{G_F}{\sqrt{2}}V_{uD}\bar{\nu}_\tau\gamma_\mu(1-\gamma_5)\tau \bar{D}\gamma^\mu(1-\gamma_5) u\,,
\end{equation}
where $G_F$ denotes the Fermi constant, $V_{uD}$ corresponds to the CKM matrix elements, with the down-type quarks $D=d$ and $s$. The amplitude of the two-pseudoscalar boson tau decay process, i.e., $\tau(p_\tau) \to P_1 (p_1) P_2 (p_2) \nu_\tau (p_\nu)$, can be then written as 
\begin{equation}\label{eq.defmt}
 \mM= \frac{G_FV_{uD}}{\sqrt2} \bar{u}(p_\nu) \gamma_\mu(1-\gamma_5)u(p_\tau) H^\mu\,,
\end{equation}
where the hadronic matrix element of the current is usually parameterized by
\begin{equation}\label{eq.defhmu}
H^\mu\equiv\langle P_1 P_2| \bar{D} \gamma^\mu u | 0 \rangle= \left[ (p_{2} - p_{1})^\mu - \frac{\Delta_{P_2P_1}}{s} q^\mu \right] F^{P_1P_2}_{+}(s) +  \frac{\Delta_{Du}}{s} q^\mu \widehat{F}^{P_1P_2}_{0}(s)\,, 
\end{equation}
with 
\begin{equation}
\Delta_{P_2P_1}=m_{P_2}^2-m_{P_1}^2\,, \qquad \Delta_{Du}=B_0(m_{D}-m_u)\,, \qquad q_\mu =(p_1+p_2)_\mu, \qquad s=q^2. 
\end{equation}
Here $m_{P_i}$ denotes the mass of the pseudoscalar boson $P_i$, and $m_{D=d,s}$ and $m_u$ are the light-flavor quark masses. The quantity $B_0$ relates with the nonperturbative quark condensate via $\langle0| \bar{q}_i q_j |0\rangle=-F^2 B_0\delta_{ij}$, with $F$ the pion decay constant in chiral limit. In this parameterization, $F^{P_1P_2}_{+}(s)$ and $\widehat{F}^{P_1P_2}_{0}(s)$ represent the vector and scalar form factors, respectively. The finitude feature of the matrix elements in Eq.~\eqref{eq.defhmu} at $s=0$ requires the following normalization condition 
\begin{eqnarray}\label{eq.fvfsat0}
F_{+}^{P_1P_2}(0)=\frac{\Delta_{Du}}{\Delta_{P_2P_1}} \widehat{F}_{0}^{P_1P_2}(0)\,. 
\end{eqnarray}

After the evaluation of $|\mM|^2$ in Eq.~\eqref{eq.defmt} by taking the spin average/sum of the $\tau$/$\nu_\tau$, one can acquire the conventional form of the differential decay width of the $\tau^-\to (P_1 P_2)^-\nu_\tau$ process 
\begin{equation}\label{eq.dfw}
\begin{aligned}
&\frac{d^2\Gamma_{\tau \to P_1 P_2 \nu_\tau}}{d\sqrt{s}d \cos \alpha}= \frac{G_F^2 \, \left| V_{uD} \right|^2 \, S_{\text{EW}} \left( m_\tau^2 - s \right)^2  }{128\pi^3 } \,  \bigg\{ \frac{\left| \widehat{F}_0^{P_1P_2}(s) \right|^2 \, \Delta_{Du}^2 \, q_{P_1P_2}(s)}{\, m_\tau \, s^2} \\
&   + \frac{ 4\Delta_{Du} \, q_{P_1P_2}^2(s) \, \cos{\alpha} \, \Re \left[  F_+^{P_1P_2}(s)  \, \widehat{F}_0^{P_1P_2*}(s) \right]}{ \, m_\tau \, s^{3/2}}  + \frac{4\left| F_+^{P_1P_2}(s) \right|^2 \, q_{P_1P_2}^3(s) \, \left( m_\tau^2 \, \cos{\alpha}^2 + s \, \sin{\alpha}^2 \right)}{ \, m_\tau^3 \, s} \bigg\}\,.
\end{aligned}
\end{equation}
Here we adopt the short-distance electroweak radiative correction $S_{\text{EW}} =1.0201$ from Ref.~\cite{Erler:2002mv}. It consists of (i) the combined logarithmically enhanced electroweak and short-distance QCD corrections, by performing the next-to-leading order resummation of the logarithms through renormalization group equation; 
(ii) the remaining non-logarithmically enhanced electroweak corrections.
The quantity $\alpha$ in Eq.~\eqref{eq.dfw} denotes the angle between the momenta of one of the two pseudoscalar bosons and the $\tau$ lepton in the $P_1P_2$ rest frame, and the three-momentum of the pseudoscalar bosons in the latter frame is given by 
\begin{eqnarray}
q_{P_1P_2}(s) = \frac{\sqrt{s^2 - 2s \Sigma_{P_1P_2} + \Delta_{P_1P_2}^2}}{2\sqrt{s}},\qquad \Sigma_{P_1P_2} = m_{P_1}^2 + m_{P_2}^2\,.  
\end{eqnarray}

To further integrate out the angle $\alpha$ in Eq.~\eqref{eq.dfw}, we obtain the differential decay width with respect to $\sqrt{s}$, i.e., the energy of the $P_1P_2$ system in the center of mass frame, 
\begin{equation}\label{dGammapi}
\begin{aligned}
\frac{d\Gamma_{\tau \to P_1 P_2 \nu_\tau}}{d\sqrt{s}} =& \frac{G_F^2 M_\tau^3}{48 \pi^3 s} S_\text{EW} \left| V_{uD} \right|^2 \left( 1 - \frac{s}{M_\tau^2} \right)^2 \\& 
\left\{
\left( 1 + \frac{2s}{M_\tau^2} \right) q_{P_1P_2}^3(s) \left| F_{+}^{P_1P_2}(s) \right|^2
+ \frac{3 \Delta_{Du}^2}{4s} q_{P_1P_2}(s) \left| \widehat{F}_{0}^{P_1P_2}(s) \right|^2
\right\}\,.
\end{aligned}
\end{equation}

Alternatively, one can also construct another important type of differential distribution, viz. the so-called forward-backward (FB) asymmetry, 
\begin{equation}\label{eq.afb}
\begin{aligned}
A_{FB}(s) =& \frac{\int_0^1 d\cos \alpha \frac{d^2\Gamma_{\tau \to P_1 P_2 \nu_\tau}}{d\sqrt{s}d\cos \alpha} - \int_{-1}^0 d\cos \alpha \frac{d^2\Gamma_{\tau \to P_1 P_2 \nu_\tau}}{d\sqrt{s}d\cos \alpha}}{\int_0^1 d\cos \alpha \frac{d^2\Gamma_{\tau \to P_1 P_2 \nu_\tau}}{d\sqrt{s}d\cos \alpha} + \int_{-1}^0 d\cos \alpha \frac{d^2\Gamma_{\tau \to P_1 P_2 \nu_\tau}}{d\sqrt{s}d\cos \alpha}}
\\ 
=& \frac{ \Delta_{Du} \, q_{P_1P_2}(s) \, \Re \left[  F_+^{P_1P_2}(s)  \, \widehat{F}_0^{P_1P_2*}(s) \right]}{\frac{2\sqrt{s}}{3}\left( 1 + \frac{2s}{M_\tau^2} \right) q_{P_1P_2}^2(s) \left| F_{+}^{P_1P_2}(s) \right|^2
+ \frac{ \Delta_{Du}^2}{2\sqrt{s}}  \left| \widehat{F}_{0}^{P_1P_2}(s) \right|^2
}\,,
\end{aligned}
\end{equation}
which probes the interference term of the vector and scalar form factors that is however absent in the differential decay width expression in Eq.~\eqref{dGammapi}. It is noted that the FB asymmetry can play important roles in establishing the CP violation in hadronic $\tau$ decays~\cite{Chen:2020uxi,Chen:2021udz,Aguilar:2024ybr}. 

From the above discussions, it is clear that under the assumption of SM structure of charged currents the $\tau \to P_1 P_2 \nu_\tau$ processes are totally determined by the two form factors $F_{+}^{P_1P_2}(s)$ and $\widehat{F}_{0}^{P_1P_2}(s)$, both of which will be calculated within $\rxt$ in the following section.

\section{Calculation of the hadronic form factors}\label{sec.calcff}

\subsection{Pertinent resonance chiral Lagrangians}

Since the relevant energy region of the form factors $F^{P_1P_2}_{+}(s)$ and $\widehat{F}^{P_1P_2}_{0}(s)$ spans from the two-boson threshold up to $m_\tau$, apart from the contact meson interactions that are dictated by the chiral symmetry in the low energy region, the hadronic meson resonances play decisive roles in the intermediate energy region. In practice, it is convenient to derive the vector form factor $F^{P_1P_2}_{+}(s)$ via the matrix elements of the $\bar{D}\gamma^\mu u$ current and the scalar form factor $\widehat{F}^{P_1P_2}_{0}(s)$ via the matrix elements of $\bar{D} u$.
$\rxt$ has proven to be able to provide a reliable and sophisticated theoretical framework to calculate pertinent form factors appearing in the $\tau$ decays~\cite{Arteaga:2022xxy,Gonzalez-Solis:2019lze,Escribano:2016ntp,Escribano:2013bca,Sanz-Cillero:2017fvr,Jamin:2006tk,Guerrero:1997ku,Chen:2022nxm,Guo:2008sh,Guo:2010dv}. 
To make a self-consistent evaluation, we recapitulate the calculation of the relevant two-meson form factors to the $\tau$ decays in $\rxt$, though they have been individually addressed in previous works, such as those in Refs.~\cite{Guerrero:1997ku,Escribano:2016ntp,Escribano:2013bca,Jamin:2006tk}. As a novelty, we calculate the axion related form factors appearing in the $\tau^-\to \pi^-(K^-) a\nu_\tau$ processes, and also update the expressions involving the $\eta$ and $\eta'$ based on the newly determined $\pi^0$-$\eta$-$\eta'$-$a$ mixing formulas in Refs.~\cite{Gao:2022xqz,Gao:2024vkw}, where the model-independent axion interaction operator, viz., $\alpha_s a G\tilde{G}/(8\pi f_a)$, is introduced to describe the axion-meson system.  

In this work, it is reiterated that we will also stick to the model-independent axion interaction $aG\tilde{G}/f_a$. In particular, we do not consider the axion-lepton interaction vertex throughout, while this latter type of interaction has been recently explored in the tau decays in Ref.~\cite{Ema:2025bww}. It is demonstrated in Ref.~\cite{Gao:2022xqz} that the $U(3)$ axion chiral theory that explicitly includes the QCD $U_A(1)$ anomaly effect provides a viable framework to simultaneously incorporate the axion, $\pi$, $K$, $\eta$ and $\eta'$ states. The leading order (LO) $U(3)$ Lagrangian reads 
\begin{eqnarray}\label{eq.laglo}
\mathcal{L}^{\rm LO}= \frac{F^2}{4}\langle u_\mu u^\mu \rangle+
\frac{F^2}{4}\langle \chi_+ \rangle
+ \frac{F^2}{12}M_0^2 X^2 \,,
\end{eqnarray}
where $\langle ...\rangle$ stands for the trace over the light-quark flavor space, and the last term corresponds to the explicit $U_A(1)$ anomaly term, responsible for the large mass $M_0$ of the singlet $\eta_0$. The associated chiral building operators are defined as:  
\begin{eqnarray}
& u_\mu = i u^\dagger  D_\mu U u^\dagger \, , \qquad \chi_\pm  = u^\dagger  \chi u^\dagger  \pm  u \chi^\dagger  u \,, \qquad U =  u^2 = e^{i\frac{ \sqrt2\Phi}{ F}}\,,\\ &\qquad D_\mu U =\partial_\mu U-i(v_\mu+a_\mu) U +iU (v_\mu-a_\mu)  \,, \qquad \chi = 2B_0(s+ip) \,,\qquad
\end{eqnarray} 
where $v_\mu$, $a_\mu$, $s$, $p$ are the chiral external sources with types of vector, axial-vector, scalar and pseudoscalar, respectively. The matrix representation of the pseudo-Nambu Goldstone bosons (pNGBs) is given by 
\begin{equation}\label{phi1}
\Phi \,=\, \left( \begin{array}{ccc}
\frac{1}{\sqrt{2}} \pi^0+\frac{1}{\sqrt{6}}\eta_8+\frac{1}{\sqrt{3}} \eta_0 & \pi^+ & K^+ \\ \pi^- &
\frac{-1}{\sqrt{2}} \pi^0+\frac{1}{\sqrt{6}}\eta_8+\frac{1}{\sqrt{3}} \eta_0   & K^0 \\  K^- & \overline{K}^0 &
\frac{-2}{\sqrt{6}}\eta_8+\frac{1}{\sqrt{3}} \eta_0
\end{array} \right)\,.
\end{equation}
In $U(3)$ chiral theory, the axion field can be introduced into the chiral Lagrangian via~\cite{Gao:2022xqz,Gao:2024vkw} 
\begin{eqnarray}
X= \log{(\det U)} - i\frac{a}{f_a}\,,  
\end{eqnarray}
without performing the axial transformation of the quark fields as commonly used in chiral studies~\cite{GrillidiCortona:2015jxo,DiLuzio:2020wdo}. 

Next, we provide the relevant resonance chiral Lagrangians to our study. According to the seminal work of $\rxt$~\cite{Ecker:1988te}, the minimal interacting Lagrangians involving vector and scalar resonances are 
\begin{equation}\label{eq.lagv}
\mathcal{L}_V = \frac{F_V}{2\sqrt{2}} \langle V_{\mu\nu} f_+^{\mu\nu} \rangle + \frac{iG_V}{\sqrt{2}} \langle V_{\mu\nu} u^\mu u^\nu \rangle  \, ,
\end{equation}
\begin{equation}\label{eq.lags}
\mathcal{L}_S = c_d \langle S u_\mu u^\mu \rangle + c_m \langle S \chi_+
\rangle \, ,
\end{equation}
where $V_{\mu\nu}$ is the vector resonance nonet in the anti-symmetric tensor form, $S$ stands for the scalar resonance nonet and the large $N_C$ relations are imposed to both the vector ($F_V,G_V$) and scalar ($c_d,c_m$) resonance couplings. Since we investigate the two-boson interactions from their thresholds up to $m_\tau$ in the current study, several excited multiplets of hadron resonances, instead of just the low-lying ones whose couplings are denoted by the unprimed symbols, like $F_V, G_V, c_{m,d}$, can play relevant roles. We will use the same forms of interacting operators, but with the obvious replacement of resonance couplings by introducing primes, such as $F_V'$, $G_V'$ and $c_{m,d}'$, to include the contributions from the first excited hadron resonances. We also introduce double-prime couplings, such as $F_V''$ and $G_V''$, to account for the contributions from the second excited vector resonance states. The scalar resonance masses in the chiral limit will be denoted by $M_S$ and $M_{S'}$. 
Under the resonance saturation assumption at large $N_C$, the heavy hadron resonances give the dominant contributions to the next-to-leading order (NLO) chiral low energy constants (LECs), which should not be included any more in the $\rxt$ Lagrangians~\cite{Ecker:1988te}. Otherwise, there will be double counting issues. 

\begin{figure}[h]
    \centering
        \includegraphics[width=\linewidth]{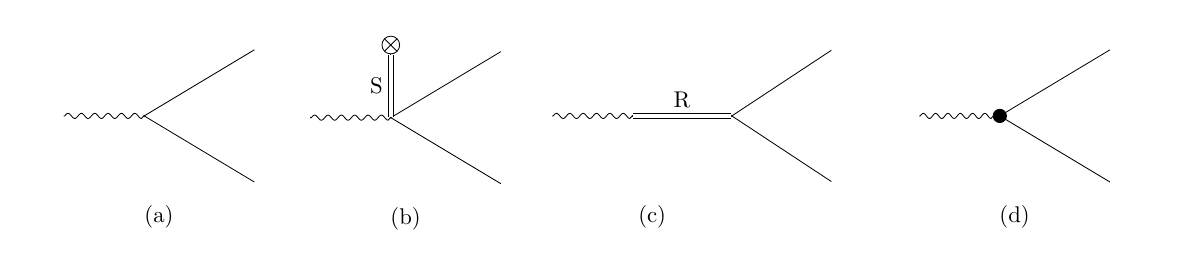}
    
    \caption{ Feynman diagrams for the two-boson form factors. The wiggly lines denote the vector (scalar) external sources for the vector (scalar) form factors. The double lines correspond the resonance fields. The circle cross symbol connected to the scalar resonance stands for the vacuum, which is caused by the tadpole contribution from the $c_m$ term in Eq.~\eqref{eq.lags}. The type of the resonance $R$ in diagram (c) is the same as the vector or scalar external source. The diagram (d) arises from the two local NLO operators in Eq.~\eqref{eq.laglam}.}
    \label{fig.feyndiag}
\end{figure}

\begin{figure}[h]
    \centering
        \includegraphics[width=\linewidth]{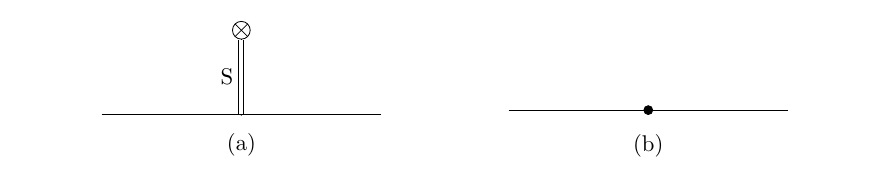}
    
    \caption{Feynman diagrams for the pseudoscalar self-energy. See Fig.~\ref{fig.feyndiag} for the meaning of different symbols. }
    \label{fig.selfenergy}
\end{figure}

However, in the $U(3)$ case, two additional local NLO operators, namely  
\begin{equation}\label{eq.laglam}
\mathcal{L}_{\Lambda}^{\rm NLO, U(3)}= -\frac{F^2\, \Lambda_1}{12}   D^\mu X D_\mu X  -\frac{F^2\, \Lambda_2}{12} X \langle \chi_- \rangle\,,
\end{equation}
can not be accounted for by the resonance exchanges at large $N_C$. Therefore we explicitly take them into account in our calculation, together with the resonance Lagrangians in Eqs.~\eqref{eq.lagv} and \eqref{eq.lags}, and also those from highly excited multiplets. The Feynman diagrams that contribute to the corresponding form factors are illustrated in Fig.~\ref{fig.feyndiag}. At the same time, we also need to compute the wave renormalization constants and masses of the pNGBs via the self-energy diagrams, as shown in Fig.~\ref{fig.selfenergy}

The neutral pNGBs, i.e., $\pi^0\,,\eta_8\,,\eta_0$, and the axion $a$ will get mixed, according to the chiral Lagrangians in Eqs.~\eqref{eq.laglo}, \eqref{eq.lags} and \eqref{eq.laglam}. Before calculating any physical quantities, such as the form factors and scattering amplitudes, one should first perform the field redefinitions to eliminate their mixing in order to get the physical states. In Refs.~\cite{Gao:2022xqz,Gao:2024vkw}, the mixing pattern among $\pi^0\,,\eta_8\,,\eta_0\,,a$ has been worked out within the $U(3)$ chiral perturbation theory by explicitly incorporating the local NLO LECs in the $\delta$ counting scheme~\cite{Kaiser:2000gs}. We can directly take the results of Refs.~\cite{Gao:2022xqz,Gao:2024vkw} to handle the mixing in the present work. The mixing formula between the bases of physical states (denoted by $\hat{\phi}$) and the flavor states can be written as
\begin{eqnarray}\label{eq.mixnlof08}
 {\tiny  \left( \begin{array}{c}
\hat{\pi}^0 \\   \hat{\eta}  \\  \hat{\eta}' \\ \hat{a} 
\end{array} \right) 
=
\left( \begin{array}{cccc}
1 + z_{11}      & c_\theta (-v_{12}+z_{12}) +s_\theta (-v_{13} + z_{13}) & -s_\theta (-v_{12}+z_{12}) +c_\theta (-v_{13} + z_{13}) & -v_{14}+z_{14} \\ 
v_{12} + z_{21} & c_\theta (1+ z_{22}) +s_\theta  ( z_{23} -v_{23}) & -s_\theta (1+ z_{22}) +c_\theta  (z_{23}-v_{23})    & -v_{24}+z_{24}  \\ 
v_{13} + z_{31} & c_\theta  ( z_{32} + v_{23}) +s_\theta (1+ z_{33})   & -s_\theta  (z_{32} + v_{23}) + c_\theta (1+ z_{33})    & -v_{34}+z_{34}  \\ 
v_{41}+z_{41}   & c_\theta (v_{42}+z_{42}) +s_\theta (v_{43}+z_{43})   & -s_\theta (v_{42}+z_{42})+ c_\theta (v_{43}+z_{43}) & 1+ v_{44}+ z_{44} 
\end{array} \right)  
\left( \begin{array}{c}
\pi^0 \\  \eta_8 \\  \eta_0 \\ a
\end{array} \right) }\,, 
\end{eqnarray}
where $v_{ij}$ and $z_{ij}$ correspond to the LO and NLO contributions, respectively. In later discussions, we designate $\pi^0,\eta,\eta,a$ as the physical states for simplicity. $c_\theta$ and $s_\theta$ are short for $\cos\theta$ and $\sin\theta$, with $\theta$ the LO $\eta$-$\eta'$ mixing angle. The NLO coefficients $z_{ij}$ encode two parts denoted as $x_{ij}$ and $y_{ij}$ in Ref.~\cite{Gao:2022xqz,Gao:2024vkw}, where $x_{ij}$ are responsible for the kinematic mixing and $y_{ij}$ are introduced to address the mass mixing at NLO. Though the expressions of $x_{ij}$ and $y_{ij}$ are rather lengthy, they only depend on four NLO LECs, namely $L_5,\, L_8,\, \Lambda_1$ and $\Lambda_2$. To make a consistent calculation, we will use the resonance saturation to estimate $L_5$ and $L_8$ in this work, which are given by $L_5=c_dc_m/M_S^2+c_d'c_m'/M_{S'}^2$ and $L_8=c_m^2/(2M_S^2)+c_m'^2/(2M_{S'}^2)$~\cite{Ecker:1988te}. The effects of the pseudoscalar resonances are neglected for the latter quantity. The values of $L_5$ and $L_8$ by taking the resonance parameters discussed later in our study are compatible with those obtained from the comprehensive Bayesian method in Ref.~\cite{Pan:2023qja}.
The explicit expressions of $v_{ij}$, $x_{ij}$, $y_{ij}$ and $z_{ij}$ are given in Ref.~\cite{Gao:2024vkw}, and we do not repeat them here.  

\subsection{Strangeness conserving form factors}

The decays of $\tau^-\to\pi^-(\pi^0,\eta,\eta',a)\nu_\tau$ are driven by the strangeness conserving currents, belonging to the Cabibbo allowed processes. In this case, we need  to evaluate the matrix elements of the strangeness conserving vector current $\langle P_1 P_2|\bar{d}\gamma_{\mu} u |0\rangle$ to acquire the corresponding vector form factors. Regarding the scalar form factors, the proper object to extract them is $\langle P_1 P_2| (m_D-m_u)\bar{D}u |0\rangle$, since the latter quantity corresponds to the divergence of the vector current, i.e., $\langle P_1 P_2| \partial^\mu(\bar{D}\gamma_\mu u) |0\rangle=i\langle P_1 P_2| (m_D-m_u)\bar{D}u |0\rangle$, and is renormalization group invariant~\cite{Jamin:2001zr,Jamin:2001zq}. 

In practical calculation, we will substitute the light quark masses $m_{D=d,s}$ and $m_u$ with the physical masses of $\pi$ and $K$ evaluated in $\rxt$. For the strangeness conserving situation, the isospin breaking quantity $m_d-m_u$ enters in the definition of the scalar form factors. According to the Dashen's theorem~\cite{Dashen:1969eg}, the electromagnetic (EM) corrections to the pion and kaon masses are equal, i.e., $\delta_{m_{K^0}^2-m_{K^{+}}^2}^{\rm EM} = \delta_{m_{\pi^0}^2-m_{\pi^{+}}^2}^{\rm EM}$. By further taking into account the resonance contribution to the kaon masses, up to NLO accuracy one can rewrite the light-quark mass $m_d-m_u$ in terms of the physical meson masses as  
\begin{equation}\label{eq.deldu}
\Delta_{du}=B_0(m_d - m_u) = \Delta_{du}^{\text{Phy}} \left\{ 1 + \frac{m_K^2}{F^2} \left[\frac{16 c_m(c_d - c_m) }{ M_S^2}+\frac{16 c_m'(c_d' - c_m') }{ M_{S'}^2} \right] \right\} \,,
\end{equation}
where $\Delta_{du}^{\text{Phy}} = m_{K^0}^2 -m_{K^{+}}^2 - (m_{\pi^0}^2- m_{\pi^{+}}^2 )$, $m_K^2$ inside the curly brackets denotes the isospin-averaged kaon mass squared. In this work we include the effects of isospin breaking (IB) up to the linear order for the scalar component of the hadronic matrix element of $\bar{d}\gamma^\mu u$. In this case,  it is justified to use the physical isospin averaged mass squared inside the curly brackets of Eq.~\eqref{eq.deldu}. The strangeness conserving scalar form factor is then defined as 
\begin{equation}
(m_d - m_u) \langle \pi^- P | \bar{d} u | 0 \rangle
= \Delta_{du}^{\text{Phy}}  F_0^{\pi^- P}(s)\,,  
\end{equation}
with $P=\pi^0,\eta,\eta',a$. Comparing with the definition in Eq.~\eqref{eq.defhmu}, the product of $\Delta_{du}\widehat{F}_0^{\pi^- P}(s)$ in fact equals to $\Delta_{du}^{\text{Phy}}  F_0^{\pi^- P}(s)$, which leads to 
\begin{equation}
 \widehat{F}_0^{\pi^- P}(s)=\frac{\Delta_{du}^{\text{Phy}} }{\Delta_{du}} F_0^{\pi^- P}(s)\,. 
\end{equation}
Correspondingly, the matrix elements of the strangeness conserving vector current in Eq.~\eqref{eq.defhmu} can be recast as 
\begin{equation}\label{eq.defhmuphyud}
\langle \pi^- P | \bar{d} \gamma^\mu u | 0 \rangle = \left[ (p_{P} - p_{\pi})^\mu - \frac{\Delta_{P \pi }}{s} q^\mu \right] F^{\pi^- P}_{+}(s) +  \frac{\Delta^{\text{Phy}}_{du}}{s} q^\mu F^{\pi^- P}_{0}(s)\,. 
\end{equation}
It is pointed out that many other works, e.g., Refs.~\cite{Cheng:2017pcq,Faustov:2019mqr}, parameterize the scalar component of the matrix element $\langle P_1 P_2| \bar{D} \gamma^\mu u | 0 \rangle$ as $\Delta_{P_1P_2}\bar{F}_0(s)q^\mu/s$, with $\Delta_{P_1P_2}=m_{P_1}^2-m_{P_2}^2$. Since $\Delta_{P_1P_2}$ is not always proportional to $\Delta_{Du}$, this indicates that the scalar form factor $\bar{F}_0(s)$ would compromise the flavor symmetry breaking factor in such cases. While, in our convention the factor proportional to $m_D-m_u$ is factored out and the scalar form factors $\hat{F}_0(s)$ and $F_0(s)$ do not contain such flavor symmetry breaking terms.

For the decay channel of $\tau^- \to \pi^-\pi^0\nu_\tau$, compared to the vector form factor, the contribution from the scalar form factor is much suppressed by the isospin breaking factor. Therefore we will neglect the tiny effect from the scalar form factor. The vector form factor appearing in this decay channel takes the form 
\begin{equation}\label{eq.ffpipi}
\begin{aligned}
F^{\pi^-\pi^0}_+(s)  = & -\frac{\sqrt{2}}{F^2} G_{\rm{LO}+\rm{\rho\, Ex}}(s)\,, 
\end{aligned}
\end{equation}
with 
\begin{equation}\label{eq.grhoex}
\begin{aligned}
G_{\rm{LO}+\rm{\rho\,Ex}}(s)  = &  \frac{ G_V F_V s+F^2(M_\rho^2-s)}{M_\rho^2 - s -iM_\rho\Gamma_\rho(s)}  -  \frac{G_V' F_V' s}{M_{\rho'}^2 - s -iM_{\rho'}\Gamma_{\rho'}(s)}-\frac{ G_V'' F_V'' s}{M_{\rho''}^2 - s -iM_{\rho''}\Gamma_{\rho''}(s)} \,, 
\end{aligned}
\end{equation}
where we follow the recipe of $\rxt$ Lagrangian in Eq.~\eqref{eq.lagv} to also introduce another two excited resonances $\rho'$/$\rho(1450)$ and $\rho''$/$\rho(1700)$, in addition to the ground state $\rho$/$\rho(770)$. We point out the subtle issue about the minus signs in front of the $\rho'$ and $\rho''$ terms in Eq.~\eqref{eq.grhoex}. In fact, it is common to introduce some phase parameters for different resonance states to describe the $\tau\to\pi\pi\nu_\tau$ process in many existing works~\cite{Kuhn:1990ad,Schneider:2012ez,Shekhovtsova:2012ra,GomezDumm:2013sib,Gonzalez-Solis:2019iod}. We take advantage of these previous studies to fix minus signs for the $\rho'$ and $\rho''$ a posteriori, allowing us to make a good description of the precise $\pi\pi$ spectra from Belle~\cite{Belle:2008xpe}. In this way, the number of free parameters is reduced and the fit tends to be stable.

We further extend the discussion to the decay channel of $\tau^- \rightarrow \pi^-(\eta,\eta',a)\nu_\tau$. Unlike the $\tau^- \to \pi^-\pi^0\nu_\tau$ channel, there is no relative suppression between the vector and scalar form factors, although an overall isospin breaking factor appears in each of the three processes of $\tau^- \rightarrow \pi^-(\eta,\eta',a)\nu_\tau$. As a result, one should consider both contributions from the vector and scalar form factors to these processes. 
To make a more realistic description of the differential decay widths, we also introduce two scalar nonet resonances. We use $c_m$ and $c_d$ to denote the resonance parameters for the low-lying scalar multiplet containing the relevant $a_0(980)$ (simply denoted as $a_0$) and $K_0^*(800)$ (denoted as $K_0^*$) in this work, and designate the primed resonance parameters $c_m^\prime$ and $c_d^\prime$ for the second multiplet of scalar resonances containing $a_0(1450)$ (denoted as $a_0'$) and $K_0^*(1430)$ (denoted as $K_0^{*'}$) that are pertinent to the present study. The corresponding vector form factors take the form 
\begin{equation}\label{eq.ffvpieta}
\begin{aligned}
F^{\pi^-\eta}_+(s)=&-\frac{\sqrt{2}v_{12}}{F^2} G_{\rm{LO}+\rm{\rho\,Ex}}(s)   -\sqrt{2} \left(y_{12} - v_{13} y_{23}^{(0)}\right)\,, 
\end{aligned}
\end{equation}
\begin{equation}
\begin{aligned}
F^{\pi^-\eta'}_+(s)=&-\frac{\sqrt{2}v_{13}}{F^2} G_{\rm{LO}+\rm{\rho\,Ex}}(s) -\sqrt{2} \left(y_{13}  + v_{12} y_{23}^{(0)}\right)\,,
\end{aligned}
\end{equation}
\begin{equation}\label{VFFpia}
\begin{aligned}
F^{\pi^- a}_+(s)=&-\frac{\sqrt{2} v_{41}}{F^2}G_{\rm{LO}+\rm{\rho\,Ex}}(s)-\sqrt{2}\left( y_{14} + v_{12}\, y_{24}^{(0)} + v_{13}\, y_{34}^{(0)} \right) \,, 
\end{aligned}
\end{equation}
and the scalar form factors are given by
\begin{equation}
\begin{aligned}
F^{\pi^-\eta}_0(s) = &\sqrt{\frac{2}{3}}(c_\theta  - \sqrt{2} s_\theta )  +\frac{1}{\sqrt{3}}(\Lambda_1 - 2 \Lambda_2) s_\theta - \frac{1}{\sqrt{3}}y_{23} (2 c_\theta  +  \sqrt{2} s_\theta )+ 4\sqrt{\frac{2}{3}}\frac{c_\theta  - \sqrt{2} s_\theta}{F^2} \bigg\{  \\ & \bigg[\frac{c_m(c_m - c_d)2m_\pi^2 + c_mc_d(s + m_\pi^2 - m_{\eta}^2)}{ M_{a_0}^2 - s -i M_{a_0} \Gamma_{a_0}(s)}  - \frac{2c_m(c_m - c_d)(2m_K^2 - m_\pi^2)}{  M_{S}^2} \bigg] \\& 
+ \bigg[ c_{m,d},M_{a_0},\Gamma_{a_0},M_S \to c_{m,d}',M_{a_0'},\Gamma_{a_0'},M_{S'}    \bigg]
\bigg\}   \,,
\end{aligned}
\end{equation}
\begin{equation}
\begin{aligned}
F^{\pi^-\eta'}_0(s) =& \sqrt{\frac{2}{3}}( \sqrt{2} c_\theta  + s_\theta ) - \frac{1}{\sqrt{3}}(\Lambda_1 - 2 \Lambda_2)  c_\theta  + \frac{1}{\sqrt{3}}y_{23} (\sqrt{2} c_\theta  - 2 s_\theta )+ 4\sqrt{\frac{2}{3}}\frac{ \sqrt{2} c_\theta  + s_\theta}{F^2} \bigg\{ \\ &\bigg[ \frac{c_m(c_m - c_d)2m_\pi^2 + c_m c_d(s + m_\pi^2 - m_{\eta'}^2)}{ M_{a_0}^2 - s-i M_{a_0} \Gamma_{a_0}(s)}  - \frac{2c_m(c_m - c_d) 
 (2m_K^2 - m_\pi^2)}{ M_{S}^2} \bigg] \\ &
+ \bigg[ c_{m,d},M_{a_0},\Gamma_{a_0},M_{S} \to c_{m,d}',M_{a_0'},\Gamma_{a_0'},M_{S'}     \bigg]\bigg\} \,,
\end{aligned}
\end{equation}
\begin{equation}\label{eq.SFFpia}
\begin{aligned}
F^{\pi^- a}_0(s)=&\frac{ (\sqrt{2} c_\theta -2 s_\theta )}{\sqrt{3}} v_{24}^{(0)} + \frac{ (2 c_\theta  +\sqrt{2} s_\theta )}{\sqrt{3}}v_{34}^{(0)} -\frac{2}{\sqrt{3}} \left( s_\theta \, v_{24}^{(0)} - c_\theta \, v_{34}^{(0)} + \frac{F}{\sqrt{6}f_a}\right) (\Lambda_2-\frac{1}{2} \Lambda_1)\\ & + \frac{ (\sqrt{2} c_\theta -2 s_\theta )}{\sqrt{3}} y_{24}^{(0)}  + \frac{(2 c_\theta  +\sqrt{2} s_\theta )}{\sqrt{3}} y_{34}^{(0)}+\frac{ 4(\sqrt{2} c_\theta -2 s_\theta )v_{24}^{(0)}+4(2 c_\theta  +\sqrt{2} s_\theta )v_{34}^{(0)}}{\sqrt{3}F^2} \bigg\{\\ &\bigg[\frac{  2c_m^2\, m_\pi^2 +  c_m c_d \left( s - m_a^2- m_\pi^2 \right) }{   M_{a_0}^2 - s -i M_{a_0} \Gamma_{a_0}(s)}  + \frac{2c_m(c_d-c_m)\,(2 m_K^2-m_\pi^2)}{ M_{S}^2 }  \bigg]\\ & + \bigg[ c_{m,d},M_{a_0},\Gamma_{a_0},M_{S} \to c_{m,d}',M_{a_0'},\Gamma_{a_0'},M_{S'}  \bigg]\bigg\}\,. 
\end{aligned}
\end{equation}
It is reiterated that the mixing items $v_{ij}\,$ and $y_{ij}$ appearing in above equations are explicitly given in Ref.~\cite{Gao:2024vkw}. The values of $L_5$ and $L_8$ entering in $y_{ij}$ will be estimated by the resonance saturation assumption in this work. 
The hadronic matrix element $H_\mu$ of Eq.~\eqref{eq.defhmuphyud} should remain finite as $s \to 0$, which in turn requires the following normalization condition 
\begin{eqnarray}\label{normalization}
    F_{+}^{\pi^{-}(\eta,\eta',a)}(0)=\frac{\Delta_{du}^{\text{Phy}} }{\Delta_{(\eta,\eta',a) \pi }} F_{0}^{\pi^{-} (\eta,\eta',a)}(0)\,. 
\end{eqnarray}
By carefully expanding the right hand side of the above equation up to the NLO in the $U(3)$ $\delta$ counting scheme, i.e., the simultaneous expansions of momentum, quark mass and $1/N_C$, and combining the results of Ref.~\cite{Gao:2024vkw}, we have explicitly verified that the vector and scalar form factors in Eqs.~\eqref{eq.ffvpieta}-\eqref{eq.SFFpia} indeed satisfy the relations in Eq.~\eqref{normalization}. 
We point out several subtleties about the calculation of scalar form factors. For the scalar hadron exchanges in the diagram (b) of Fig.~\ref{fig.feyndiag}, i.e., with the vertices between the unflavored scalar hadrons and the vacuum, we will take the scalar masses $M_S$ and $M_{S'}$ in the chiral and large $N_C$ limits. While for the scalar resonance exchanges in the diagram (c) of Fig.~\ref{fig.feyndiag}, we use physical parameters for different intermediate scalar resonances in the propagators, including the masses and finite widths, which are expected to offer a more reliable description of the two-pseudoscalar boson spectra. It is noted that there are other more rigorous ways to include the scalar resonances in the form factors~\cite{Zheng:2013dda,VonDetten:2021rax}, while we stick to the Lagrangian method mentioned above, since this latter approach is more flexible to fit the experimental data and is also easier to relate the scalar dynamics in different channels, allowing us to make predictions to the channels yet to be measured.

Regarding the energy-dependent finite decay widths of the vector resonances appearing in Eq.~\eqref{eq.grhoex}, we follow Ref.~\cite{Gonzalez-Solis:2019iod} to construct their forms. The expressions for the three vector resonances $\rho$, $\rho'$ and $\rho''$ are 
\begin{equation}\label{Gammarou}
\Gamma_\rho(s) 
= \frac{M_\rho s}{96 \pi F_\pi^2} \left[ \sigma_{\pi\pi}^3(s)   + \frac{1}{2} \sigma_{KK}^3(s)   \right]\,,
\end{equation}
\begin{equation}
\Gamma_{\rho', \rho''}(s) = \Gamma_{\rho', \rho''} \frac{s}{M_{\rho', \rho''}^2} \frac{\sigma_{\pi\pi}^3(s)}{\sigma_{\pi\pi}^3(M_{\rho', \rho''}^2)} \,,
\end{equation}
with 
\begin{equation}
\sigma_{P_1P_2}(s) = \frac{2 q_{P_1P_2}(s)}{\sqrt{s}} \theta [ s - (m_{P_1} + m_{P_2})^2 ]\,,
\end{equation}
where the values of the masses and widths related to the $\rho$-type resonances will be fitted to the $\pi\pi$ vector form factor measured in the $\tau\to\pi\pi\nu_\tau$ process~\cite{Belle:2008xpe}. 

For the energy-dependent widths of the scalar resonances of $a_0(980)$ and $a_0(1430)$, we use the forms suggested in Ref.~\cite{Escribano:2016ntp}  
\begin{equation}
\Gamma_{a_0}(s) = \Gamma_{a_0}(M_{a_0}^2)\, \left( \frac{s}{M_{a_0}^2} \right)^{3/2} \frac{g_{a_0}(s)}{g_{a_0}(M_{a_0}^2)}\,,
\end{equation}
with 
\begin{equation}\label{eq.gRinfw}
\begin{aligned}
g_{a_0}(s) = & \sigma_{K^-K^0}(s) + 2\left( \frac{c_\theta  - \sqrt{2} s_\theta }{\sqrt{3}}\right)^2 \left( 1 + \frac{\Delta_{\pi^-\eta}}{s} \right)^2 \sigma_{\pi^-\eta}(s)
\\ &+ 2 \left(\frac{s_\theta  + \sqrt{2} c_\theta }{\sqrt{3}} \right)^2\left( 1 + \frac{\Delta_{\pi^-\eta'}}{s} \right)^2 \sigma_{\pi^-\eta'}(s)\,. 
\end{aligned}
\end{equation}
Since the current available data in the $\tau$ decays are not able to effectively constrain the masses and widths of the scalar resonances $a_0(980)$ and $a_0(1430)$, we will fix their parameters through reproducing the pole positions given by the Particle Data Group (PDG)~\cite{ParticleDataGroup:2024cfk}. The pole positions in the complex energy plane correspond to the zeros of $M_{R}^2-s- i M_R \Gamma_R(s)$, appearing in the denominators of the resonance exchange terms in the aforementioned form factors, on a given Riemann sheet (RS). Taking the $a_0$ resonance as an example for illustration, the expression in Eq.~\eqref{eq.gRinfw} corresponds to the first/physical RS. The second RS is obtained by reversing the sign of the $\sigma_{\pi^-\eta}(s)$ term from lightest threshold $\pi^-\eta$, while keeping the other terms unchanged. The third RS is obtained by reversing the signs of both $\sigma_{\pi^-\eta}(s)$ and $\sigma_{K^-K^0}(s)$, leaving $\sigma_{\pi^-\eta'}(s)$ untouched. The first, second and third RSs can be denoted by the notations of $(+,+,+),(-,+,+)$ and $(-,-,+)$, respectively, where $+/-$ in each entry denotes the sign of $\sigma_{P_1P_2}$ for the thresholds in an ascending order. For the $a_0(980)$,  its mass and width parameters are determined to be $M_{a_0}=1.022$~GeV and $\Gamma_{a_0}(M_{a_0})=0.118$~GeV, in order to reproduce the pole position at $\sqrt{s_{a_0}}=(0.995-i0.050)$~GeV~\cite{ParticleDataGroup:2024cfk} on the third RS. The pole position at $\sqrt{s_{a_0'}}=(1.395-i0.085)$~GeV on the third RS requires $M_{a_0'}=1.412$~GeV and $\Gamma_{a_0'}(M_{a_0'})=0.204$~GeV. Such values of the masses' and widths' parameters will be exploited in later phenomenological studies.

\subsection{Strangeness changing form factors}

The decays of $\tau^-\to K_S\pi^- \nu_\tau$ and $\tau^-\to K^-(\eta,\eta',a)\nu_\tau$ are governed by the strangeness changing currents,  belonging to the Cabibbo suppressed reactions. Following the similar procedure elaborated in the previous subsection, we now need to calculate 
$\langle P_1 P_2|\bar{s}\gamma_{\mu} u |0\rangle$ and $\langle P_1 P_2| (m_s-m_u)\bar{s}u |0\rangle$ to determine the strangeness changing vector and scalar form factors, respectively.  

Following the discussion of Eq.~\eqref{eq.deldu}, we have 
\begin{equation}
\Delta_{su}=B_0(m_s - m_u) =\Delta_{su}^{\text{Phy}} \left\{ 1 + \frac{8 (m_K^2+m_\pi^2)}{F^2}\left[ \frac{ c_m(c_d - c_m)}{M_S^2} + \frac{ c_m'(c_d' - c_m')}{M_{S'}^2}  \right]   \right\}\,,
\end{equation}
where $\Delta_{su}^{\text{Phy}}=m_K^2-m_\pi^2$, being $m_K$ and $m_\pi$ the physical isospin-averaged masses. 
The strangeness changing scalar form factor is then defined as 
\begin{equation}
(m_s - m_u) \langle (KP)^- | \bar{s} u | 0 \rangle
= \Delta_{su}^{\text{Phy}}  F_0^{(KP)^-}(s)\,,  
\end{equation}
with $P=\pi,\eta,\eta',a$. The product of $\Delta_{su}\widehat{F}_0^{(KP)^-}(s)$ in fact equals to $\Delta_{su}^{\text{Phy}}  F_0^{(KP)^-}(s)$, which leads to 
\begin{equation}
 \widehat{F}_0^{(KP)^-}(s)=\frac{\Delta_{su}^{\text{Phy}} }{\Delta_{su}} F_0^{(KP)^-}(s)\,. 
\end{equation}
Correspondingly, the matrix elements of the strangeness changing vector current in Eq.~\eqref{eq.defhmu} can be recast as 
\begin{equation}\label{eq.defhmuphyus}
\langle (KP)^- | \bar{s} \gamma^{\mu} u | 0 \rangle =  \left[ (p_{P} - p_{K})^\mu - \frac{\Delta_{PK}}{s} q^\mu \right]  F_+^{(KP)^-}(s) + \frac{\Delta_{su}^{\text{Phy}}}{s} q^\mu  F_0^{(KP)^-}(s)\,.
\end{equation}
The normalization conditions for the strangeness changing vector and scalar form factors read 
\begin{eqnarray}
F_{+}^{(KP)^-}(0)= \frac{\Delta_{su}^{\text{Phy}}}{\Delta_{PK}} F_{0}^{(KP)^-}(0)\,,
\end{eqnarray}
which guarantees the finiteness of the matrix elements in Eq.~\eqref{eq.defhmuphyus} at $s=0$.

Before presenting the vector and scalar form factors, it should be noted that for the $\tau^- \to [K_S\pi^-,K^-\eta/\eta']\nu_\tau$ channels, we disregard the isospin breaking contributions, as their effects  are tiny. However, for the $\tau^- \to K^- a \nu_\tau$ process, the isospin breaking is found to give noticeable effect, due to the relatively large contribution via the $\pi^0$-$a$ mixing, i.e., the isospin-breaking term $v_{41}$ in Ref.~\cite{Gao:2024vkw}, which is basically proportional to the factor $(m_d-m_u)/(m_d+m_u)$ when taking the vanishing axion mass. Thus, we retain the linear isospin-breaking terms as provided in Ref.~\cite{Gao:2024vkw} for the vector and scalar form factors in the $\tau^- \to K^- a \nu_\tau$ decay.
Similar to the case of strangeness conserving channels in previous subsection, we include three sets of vector resonances, i.e., $K^*(892),K^{*}(1410)$  and $K^{*}(1680)$, which are simply denoted by $K^*, K^{*'}$ and $K^{*''}$ in order, and the corresponding vector form factors are 
\begin{equation}
\begin{aligned}
F^{K_S\pi^-}_+(s)=&-\frac{1}{\sqrt{2}F F_K}G_{\rm LO+K^*\,Ex}(s)\,,
\end{aligned}
\end{equation}
\begin{equation}
\begin{aligned}
F^{K^-\eta}_+(s)=&-\sqrt{\frac{3}{2}} \frac{c_\theta}{F F_K}  G_{\rm LO+K^*\,Ex}(s)+\sqrt{\frac{3}{2}} s_\theta  y_{23}^{(0)}\,,
\end{aligned}
\end{equation}

\begin{equation}
\begin{aligned}
F^{K^-\eta'}_+(s)=&-\sqrt{\frac{3}{2}} \frac{s_\theta}{F F_K} G_{\rm LO+K^*\,Ex}(s) - \sqrt{\frac{3}{2}}  c_\theta  y_{23}^{(0)}\,,
\end{aligned}
\end{equation}

\begin{equation}\label{VFFKa}
\begin{aligned}
F^{K^-a}_+(s)=& - \frac{1}{\sqrt{2}F F_K}G_{\rm LO+K^*\,Ex}(s) \bigg( v_{41} + \sqrt{3} c_\theta  v_{42} + \sqrt{3} s_\theta  v_{43}  \bigg)  \\
& - \frac{1}{\sqrt{6}}\bigg[  \sqrt{3} y_{14} + \sqrt{3} v_{12} y_{24}^{(0)} + \sqrt{3} v_{13} y_{34}^{(0)} + 3 c_\theta  (y_{24}+v_{23} y_{34}^{(0)}) + 3 s_\theta  (y_{34}- v_{23} y_{24}^{(0)})  \bigg]\,,
\end{aligned}
\end{equation}
with 
\begin{equation}
\begin{aligned}
G_{\rm LO+K^*\,Ex}(s)=& \frac{G_V F_V s+ F F_K(M_{K^*}^2 - s)}{M_{K^*}^2 - s - iM_{K^*}\Gamma_{K^*}(s)} +  \frac{G_V' F_V' s}{M_{K^{*'}}^2 - s - iM_{K^{*'}}\Gamma_{K^{*'}}(s)} \\ &+  \frac{G_V'' F_V'' s}{M_{K^{*''}}^2 - s - iM_{K^{*''}}\Gamma_{K^{*''}}(s)} \,.
\end{aligned}
\end{equation}
It is pointed out that we have replaced one of the $F$ with $F_K$ for the resonance exchange contributions to the form factors entering the $\tau^- \to [K_S\pi^-,K^-\eta/\eta'/a]\nu_\tau$ decays. Although formally such a replacement only has influences beyond NLO accuracy, practically it can lead to noticeable effects. This is because the resonance exchanges play significant roles in the $\tau$ decays and the replacement of $F$ (estimated by $F_\pi$) by $F_K$ can cause around 20\% changes of the resonance contributions.

Two sets of scalar resonances, i.e., $K_0^*(700)$ (labeled as $K_0^{*}$) and $K_0^*(1430)$ (labeled as $K_0^{*\prime}$), will be included in the scalar form factors, whose explicit expressions are given by
\begin{equation}
\begin{aligned}
F^{K_S\pi^-}_0(s)=&\frac{1}{\sqrt{2}} + \frac{2\sqrt{2}}{FF_K}\bigg\{  \bigg[\frac{c_m^2 (m_K^2 + m_\pi^2) + c_mc_d (s - m_\pi^2 - m_K^2 )}{  M_{K_0^*}^2 - s-i M_{K_0^*} \Gamma_{K_0^*}(s)  }  - \frac{c_m(c_m-c_d) (m_K^2 + m_\pi^2) }{M_{S}^2 }\bigg] \\ & + \bigg[ c_{m,d},M_{K_0^*},\Gamma_{K_0^*},M_S \to c_{m,d}',M_{K_0^{*'}},\Gamma_{K_0^{*'}},M_{S^\prime}\bigg]\bigg\}\,,
\end{aligned}
\end{equation}

\begin{equation}
\begin{aligned}
F_0^{K^-\eta} (s)=& -\frac{1}{\sqrt{6}}(c_\theta + 2 \sqrt{2} s_\theta)+\frac{1}{\sqrt{3}}(\Lambda_1-2 \Lambda_2)s_\theta-\frac{1}{\sqrt{6}}(2  \sqrt{2} c_\theta  -s_\theta  )y_{23}^{(0)}\\& - \frac{1}{F F_K} \bigg\{ \, \bigg[ \frac{4c_\theta }{\sqrt{6}}\frac{c_m c_d (s - m_K^2 - m_{\eta}^2) + c_m^2 (5m_K^2 - 3m_\pi^2)}{M_{K_0^*}^2 - s-i M_{K_0^*} \Gamma_{K_0^*}(s)}  \\&  + \frac{8 s_\theta}{\sqrt{3}} \frac{c_m c_d (s - m_K^2 - m_{\eta}^2) + 2 c_m^2 m_K^2 }{M_{K_0^*}^2 - s-i M_{K_0^*} \Gamma_{K_0^*}(s)} + \frac{4c_\theta }{\sqrt{6}}\frac{c_m(c_m - c_d)}{ M_{S}^2} (3m_K^2 - 5m_\pi^2) \\&  - \frac{8 s_\theta}{\sqrt{3}} \frac{2 c_m(c_m - c_d) m_\pi^2}{M_{S}^2}\bigg] + \bigg[ c_{m,d},M_{K_0^*},\Gamma_{K_0^*},M_S \to c_{m,d}',M_{K_0^{*'}},\Gamma_{K_0^{*'}},M_{S'}\bigg]\bigg\} ,
\end{aligned}
\end{equation}
\begin{equation}
\begin{aligned}
F_0^{K^-\eta'}(s) = & \frac{1}{\sqrt{6}}(2\sqrt{2} c_\theta -  s_\theta )- \frac{1}{\sqrt{3}}(\Lambda_1-2 \Lambda_2)c_\theta -\frac{1}{\sqrt{6}}(2  \sqrt{2} s_\theta  + c_\theta  )y_{23}^{(0)} \\& + \frac{1}{F F_K}\bigg\{  \bigg[\frac{8 c_\theta }{\sqrt{3}}\frac{ c_m c_d (s - m_K^2 - m_{\eta'}^2) + 2 c_m^2 m_K^2 }{ M_{K_0^*}^2 - s-i M_{K_0^*} \Gamma_{K_0^*}(s)} \\&  -\frac{4s_\theta}{\sqrt{6}} \frac{ c_mc_d (s - m_K^2 - m_{\eta'}^2) + c_m^2 (5m_K^2 - 3m_\pi^2) }{M_{K_0^*}^2 - s -i M_{K_0^*} \Gamma_{K_0^*}(s)} - \frac{8 c_\theta }{\sqrt{3}}\frac{2c_m(c_m - c_d) m_\pi^2}{ M_{S}^2} \\& - \frac{4s_\theta}{\sqrt{6}} \frac{c_m(c_m - c_d)}{M_{S}^2} (3m_K^2 - 5m_\pi^2)\bigg] + \bigg[ c_{m,d},M_{K_0^*},\Gamma_{K_0^*},M_{S} \to c_{m,d}',M_{K_0^{*'}},\Gamma_{K_0^{*'}},M_{S'}\bigg] \bigg\} ,
\end{aligned}
\end{equation}

\begin{equation}\label{SFFKa}
\begin{aligned}
F^{K^-a}_0(s)=&\frac{v_{41}}{\sqrt{2}}- v_{42} ( \frac{ c_\theta }{\sqrt{6} }+ \frac{2 s_\theta }{\sqrt{3}}) -  v_{43} (-\frac{2 c_\theta }{\sqrt{3}} + \frac{s_\theta }{\sqrt{6}}) - (\frac{\sqrt{2} F }{3 f_a} +  v_{42}\frac{2 s_\theta  }{\sqrt{3}} -v_{43}\frac{2 c_\theta  }{\sqrt{3}})(\Lambda_2-\frac{1}{2}\Lambda_1) \\ &- \frac{1}{6} \bigg\{ - 3 \sqrt{2} y_{14} + 4 \sqrt{3} s_\theta  y_{24} - 3 \sqrt{2} v_{12} y_{24}^{(0)} -  \sqrt{6} s_\theta  v_{23} y_{24}^{(0)} + \sqrt{6} s_\theta  y_{34} \\ &-  3 \sqrt{2} v_{13} y_{34}^{(0)} + 4 \sqrt{3} s_\theta  v_{23} y_{34}^{(0)}  + \sqrt{3} c_\theta ( \sqrt{2} y_{24} + 4 v_{23} y_{24}^{(0)} - 4 y_{34} +  \sqrt{2} v_{23} y_{34}^{(0)})\bigg\} \\ & +\frac{1}{FF_K}  \bigg\{  \bigg[\frac{4v_{41}}{\sqrt{2}}\frac{c_m^2 (m_K^2 + m_\pi^2) + c_mc_d (s - m_a^2 - m_K^2 )}{  M_{K_0^*}^2 - s-i M_{K_0^*} \Gamma_{K_0^*}(s)  }   -\frac{4v_{41}}{\sqrt{2}}\frac{c_m(c_m-c_d) (m_K^2 + m_\pi^2) }{M_{S}^2 }  \\ & - \frac{2 (\sqrt{2} c_\theta  + 4 s_\theta )v_{42}+2 ( -4 c_\theta + \sqrt{2} s_\theta )v_{43}}{\sqrt{3}  }\frac{ c_m c_d  (s - m_a^2 - m_K^2 + \Delta_{du}^{\text{Phy}}) }{ M_{K_0^*}^2- s -i M_{K_0^*} \Gamma_{K_0^*}(s)} \\ & + \frac{16}{\sqrt{3}}\frac{(-s_\theta v_{42} + c_\theta v_{43} )c_m^2   (m_K^2 - \Delta_{du}^{\text{Phy}}) }{ M_{K_0^*}^2- s -i M_{K_0^*} \Gamma_{K_0^*}(s)}   - 2 \sqrt{\frac{2}{3}}\frac{  ( s_\theta v_{43} + c_\theta v_{42} )c_m^2   (5 m_K^2 - 3 m_\pi^2 - 2 \Delta_{du}^{\text{Phy}}) }{ M_{K_0^*}^2- s -i M_{K_0^*} \Gamma_{K_0^*}(s)}   \\ &   -  \frac{4(  c_\theta v_{42} +  s_\theta v_{43}  ) c_m(c_m-c_d)(3m_K^2-5m_\pi^2- 2\Delta_{du}^{\text{Phy}})}{\sqrt{6}   M_{S}^2} \\ &   +  \frac{8(s_\theta v_{42} - c_\theta v_{43} )c_m(c_m-c_d)(2m_\pi^2 + 2 \Delta_{du}^{\text{Phy}})}{\sqrt{3} M_{S}^2  } \bigg]\\ & + \bigg[ c_{m,d},M_{K_0^*},\Gamma_{K_0^*},M_S \to c_{m,d}',M_{K_0^{*'}},\Gamma_{K_0^{*'}},M_{S^\prime}\bigg]\bigg\}\,. 
\end{aligned}
\end{equation}

According to Ref.~\cite{Escribano:2013bca}, the expressions of the energy-dependent widths for three vector resonances $K^*, K^{*'}$ and $K^{*''}$ in the strangeness changing form factors are taken as 
\begin{equation}
\Gamma_{K^*}(s) = \Gamma_{K^*} \frac{s}{M_{K^*}^2} \frac{\sigma_{K\pi}^3(s) + c_\theta^2 \sigma_{K\eta}^3(s) + s_\theta^2 \sigma_{K\eta'}^3(s)}{\sigma_{K\pi}^3(M_{K^*}^2)}\,,
\end{equation}
\begin{equation}
\Gamma_{K^{*'(')}}(s) = \Gamma_{K^{*'(')}} \frac{s}{M_{K^{*'(')}}^2} \frac{\sigma_{K\pi}^3(s)}{\sigma_{K\pi}^3(M_{K^{*'(')}}^2)}\,.
\end{equation}
The masses' and widths' parameters for the vector strange resonances in the above equations will be fitted to the experimental data from the $\tau\to K_S\pi\nu_\tau$ and $\tau\to K\eta\nu_\tau$ processes in Refs.~\cite{Belle:2007goc,Belle:2008jjb}. 
And for the $K_0^*(700)$ and $K_0^*(1430)$ scalar resonances, their energy-dependent widths read~\cite{Escribano:2013bca}  
\begin{equation}
\Gamma_{S}(s) = \Gamma_{S}(M_{K_0^*}^2) \left( \frac{s}{M_{S}^2} \right)^{n} \frac{g_{K_0^*}(s)}{g_{K_0^*}(M_{K_0^*}^2)}\,, \qquad [n=0 :K_0^*(700) \,\,\text{and} \,\, n=\frac{3}{2}:K_0^*(1430)]\,,
\end{equation}
with 
\begin{equation}\label{GammaK0star}
\begin{aligned}
g_{K_0^*}(s) = &\frac{3}{2}\sigma_{K\pi}(s) +\frac{1}{6}\sigma_{K\eta}(s) \left[ c_\theta  \left( 1 + \frac{3\Delta_{K\pi} + \Delta_{K\eta}}{s} \right) + 2\sqrt{2}s_\theta  \left( 1 + \frac{\Delta_{K\eta}}{s} \right) \right]^2 
\\ & +\frac{4}{3}\sigma_{K\eta'}(s) \left[ c_\theta  \left( 1 + \frac{\Delta_{K\eta'}}{s} \right) - \frac{s_\theta }{2\sqrt{2}} \left( 1 + \frac{3\Delta_{K\pi} + \Delta_{K\eta'}}{s} \right) \right]^2\,. 
\end{aligned}
\end{equation}
Similar to the discussions at the end of the previous subsection, we fix the parameters of the masses and widths for $K_0^*$ and $K_0^{*'}$ appearing in the form factors by requiring their pole positions to be consistent the values given by PDG~\cite{ParticleDataGroup:2024cfk}. The results turn out to be $M_{K_0^*}=0.771$~GeV and $\Gamma_{K_0^*}(M_{K_0^*})=0.387$~GeV when the $K_0^*$ pole is located at $\sqrt{s_{K_0^*}}=(0.680-i0.300)$~GeV on the second RS. Similarly, we obtain 
$M_{K_0^{*'}}=1.454$~GeV and $\Gamma_{K_0^{*'}}(M_{K_0^{*'}})=0.233$~GeV to reproduce the $K_0^{*'}$ pole at $\sqrt{s_{K_0^{*'}}}=(1.431-i0.110)$~GeV on the third RS.

\section{Combined fit to the \texorpdfstring{$\tau^- \to (\pi^-\pi^0,K_S\pi^-,K^-\eta)\nu_\tau$}{} decays}\label{sec.fit}

In this section, we perform the combined fit to three kinds of experimental data, including the pion vector form factors $F_+^{\pi^-\pi^0}$ measured in the $\tau^- \to \pi^-\pi^0\nu_\tau$ decay in Ref.~\cite{Belle:2008xpe}, and the invariant-mass distributions of $K_S\pi^-$~\cite{Belle:2007goc} and $K^-\eta$~\cite{Belle:2008jjb} in the $\tau^- \to \nu_\tau (K_S\pi^-,K^-\eta)$ processes. 

We use the standard $\chi^2$ function in the fit to incorporate the modulus squared of the normalized pion vector form factor
\begin{equation}\label{eq.chi12}
\chi^2_1 = \sum_{i}\left( \frac{{| \tilde{F}_+^{\pi^-\pi^0}(s_i) |^2_\text{Theo}} - {| \tilde{F}_+^{\pi^-\pi^0}(s_i)|^2_\text{Exp}}}{\sigma_{| \tilde{F}_+^{\pi^-\pi^0} (s_i)|^2_\text{Exp}}}\right)^2\,,
\end{equation}
where the normalized form factor is given by $\tilde{F}_+^{\pi^-\pi^0}(s)=F_+^{\pi^-\pi^0}(s)/\sqrt{2}$ and 62 data points are taken from Ref.~\cite{Belle:2008xpe}. 
For the two-meson invariant-mass spectra, we use the following formula to perform the fit of the normalized experimental event distributions 
\begin{equation}\label{eq.numevefit}
\mathcal{N}_{P_1 P_2}(E)\equiv\frac{\mathrm{d}N_{\text{Eve}}^{\text{bin}}}{N_{\text{Eve}}^{\text{Tot}}\mathrm{d}E} = \frac{\mathrm{d}\Gamma(\tau \to P_1P_2 \nu_\tau)}{\mathrm{d}E} \frac{1}{ \Gamma_{\tau}\,\overline{B}_{\tau \to P_1P_2 \nu_\tau}} \,,
\end{equation}
where $E=\sqrt{s}$, $N_{\text{Eve}}^{\text{bin}}$ stands for the observed number of events in each bin, $N_{\text{Eve}}^{\text{Tot}}$ is the total number of events,  and $\Gamma_\tau$ corresponds to total decay width or the inverse of lifetime for $\tau$. Energy bin sizes of 11.5 MeV and 25 MeV are used for $\tau^- \to K_S\pi^- \nu_\tau $ \cite{Belle:2007goc} and $\tau^- \to K^-\eta \nu_\tau $\cite{Belle:2008jjb}, respectively, in the experimental study. For $\tau^- \to K_S\pi^- \nu_\tau $, the analysis incorporates experimental data up to bin 90 ($\sqrt{s}$ = 1.65925 $\text{GeV}$), consistent with prior analyses~\cite{Jamin:2008qg,Boito:2010me,Escribano:2014joa}. Similar as the latter references, data points corresponding to bins 5, 6, and 7 are also excluded from the $\chi^2$ minimization due to their obvious incompatibility with theoretical expectations. It is verified that the inclusion of these three data points yields a substantially increased $\chi^2$ while inducing negligible changes in the fitted parameters. For $\tau^- \to K^-\eta \nu_\tau $, the first two data points from Belle~\cite{Belle:2008jjb} are excluded as they lie significantly below the $K^-\eta$ production threshold, and data points significantly exceeding the $\tau$ mass are also excluded.
The quantity of $\overline{B}_{\tau \to P_1P_2 \nu_\tau}$ in  Eq.~\eqref{eq.numevefit} is introduced as a normalization factor in our fit, which would be identical to the branching ratio of the $\tau \to P_1P_2 \nu_\tau$ channel for the ideal description of the experimental event distributions. 
The Belle analysis in Ref.~\cite{Belle:2007goc} for the $\tau^{-} \to K_S \pi^{-} \nu_{\tau}$ decay channel yields a branching ratio of $B_{K_S\,\pi^-}^{\text{Exp}}=(4.04 \pm 0.13) \times 10^{-3}$. Regarding the $\tau^{-} \to K^{-} \eta \nu_{\tau}$ process, Belle Collaboration reports a branching fraction of $B_{K^-\eta}^{\text{Exp}}=(1.58 \pm 0.10) \times 10^{-4}$~\cite{Belle:2008jjb}. 
To circumvent the imperfect mismatch of the normalization between theoretical and experimental event distributions, we also fit the $\overline{B}_{\tau \to P_1P_2 \nu_\tau}$ quantities in Eq.~\eqref{eq.numevefit} to the aforementioned branching ratios from Belle. Thus, the $\chi^2$ function minimized in our fit for the event distributions takes the form 
\begin{equation}\label{eq.chi22}
\chi^2_2 = \sum_{P_1P_2=\substack{\,K_S \pi^-,\\ K^- \eta}}  \mathop{\sum\nolimits'}\limits_{i} \left[ \frac{\mathcal{N}^{\text{Theo}}_{P_1P_2}(E_i) - \mathcal{N}^{\text{Exp}}_{P_1P_2}(E_i)}{\sigma_{\mathcal{N},{P_1P_2}}^{\text{Exp}}(E_i)} \right]^2 +\sum_{P_1P_2=\substack{\,K_S \pi^-,\\K^- \eta}}\left( \frac{\overline{B}_{P_1P_2}^{\text{Theo}} - B_{P_1P_2}^{\text{Exp}}}{\sigma_{B_{P_1P_2}}^{\text{Exp}}} \right)^2\,,
\end{equation}
where $\sigma^{\text{Exp}}$ stand for the corresponding experimental uncertainties. The prime symbol in the summation notation denotes the exclusion of specific points from the minimization process as mentioned previously. The number of fitted data points amounts to 87 for the $K_S\pi^-$ spectrum and 31 for the $K^-\eta$ spectrum, each accompanied by their corresponding branching fractions. The total $\chi^2$ function minimized in the fit is given by the sum of those from the three channels, i.e., $\chi^2= \chi_1^2+\chi_2^2$, with $\chi_1^2$ and $\chi_2^2$ in Eqs.~\eqref{eq.chi12} and \eqref{eq.chi22}, respectively.

Next we specify the values of the parameters used in the fit. The isospin-averaged pion and kaon masses are taken as $m_{\pi}=0.137~\text{GeV}$ and $m_{K}=0.495~\text{GeV}$. Other relevant parameter values are taken from PDG~\cite{ParticleDataGroup:2024cfk}: $m_\eta=0.547~\text{GeV},\, m_{\eta'}=0.957~\text{GeV},\, G_F=1.16637 \times 10^{-5}~\text{GeV}^{-2},\,\Gamma_\tau=2.265 \times 10^{-12}~\text{GeV},\, m_\tau=1.7769~\text{GeV},\, V_{us}=0.2243,\, V_{ud}=0.9737,\,F_K=0.110~\text{GeV}$. The value of the chiral-limit quantity $F$ will be estimated by the pion decay constant $F_\pi=0.0923~\text{GeV}$. 
For the LO mass $M_0$ of the singlet $\eta_0$ in Eq.~\eqref{eq.laglo}, we take $M_0=0.820~\text{GeV}$ from Ref.~\cite{Gao:2022xqz}, which leads to the LO $\eta$-$\eta'$ mixing angle $\theta=-19.6^\circ$. 
The values of the low-energy coupling constants $\Lambda_1$ and $\Lambda_2$ in Eq.~\eqref{eq.laglam} are fixed as $\Lambda_1=-0.17,$ and $\Lambda_2=0.06$, according to Ref.~\cite{Gao:2022xqz}. 
The masses for the scalar resonances in the chiral limit are estimated by $M_S=1.0$~GeV and $M_{S'}=1.4$~GeV. 
Concerning the scalar resonance couplings, we use the values of $\tilde{c}_{d,m}$ in Ref.~\cite{Guo:2011pa} to fix $c_m=\sqrt{3}\tilde{c}_{m}=0.027$~GeV and $c_d=\sqrt{3}\tilde{c}_{d}=0.015$~GeV, and further exploit the relationship of $c_m c_d+c_m' c_d'=\frac{F^2}{4}$, dictated by the high energy behavior of the scalar form factors~\cite{Jamin:2001zq}. Thus, among the four scalar resonance couplings $c_m, c_d, c_m'$ and $c_d'$, only one of them is a free parameter. For definiteness, we choose to fit $c_m'$ in the following discussion.

Furthermore, we find that the $K^{*''}$ resonance has little impact on our fit. Thus, we fix its mass and width to be $M_{K^{*''}}=1.718\text{GeV}$ and $\Gamma_{K^{*''}}=0.320\text{GeV}$, according to PDG~\cite{ParticleDataGroup:2024cfk}.  
When allowing the width parameter of the $K^{*'}$ resonance to vary freely in the fit, we observe that it easily floats to a rather large value. However, fixing this parameter to its PDG value at $\Gamma_{K^{*'}}=0.232~\text{GeV}$ in the fit turns out to barely change the $\chi^2$. Therefore we opt to fix this width parameter in our analysis.

\begin{table}[htbp]
\centering
\renewcommand{\arraystretch}{1.5}  
\begin{tabular}{|>{\centering\arraybackslash}c|>{\centering\arraybackslash}c||>{\centering\arraybackslash}c|>{\centering\arraybackslash}c|}
\hline
$G_V\,F_V(\text{GeV}^2)\times 10^3 $ &  $10.26^{+0.01}_{-0.01}$ & $G_V'\,F_V'(\text{GeV}^2)\times 10^3$ &  $0.64^{+0.03}_{-0.02}$ \\[1ex]  

$G_V''\,F_V''(\text{GeV}^2)\times 10^3$ & $-0.94 ^{+0.05}_{-0.05}$ & $M_\rho(\text{GeV})$ & $0.7738^{+0.0003}_{-0.0003}$ \\[1ex]

$M_{\rho'}(\text{GeV})$ & $1.409 ^{+0.004}_{-0.004}$ & $\Gamma_{\rho'}(\text{GeV}) $ & $0.338^{+0.012}_{-0.010}$ \\[1ex]

$M_{\rho''}(\text{GeV})$ & $1.842^{+0.012}_{-0.013}$ & $\Gamma_{\rho''}(\text{GeV}) $ & $0.268^{+0.025}_{-0.026}$ \\[1ex]

$ c_m'(\text{GeV})$ & $ 0.053^{+0.007}_{-0.009}$ & $M_{K^{*}}(\text{GeV}) $ & $ 0.8956^{+0.0002}_{-0.0002}$ \\[1ex]

$ \Gamma_{K^{*}}(\text{GeV}) $ & $ 0.0477^{+0.0005}_{-0.0005} $ & $M_{K^{*'}}(\text{GeV}) $ & $ 1.339^{+0.009}_{-0.009}$ \\[1ex]

$ \overline{B}_{K_S\pi^-}\times 10^3$ & $3.98^{+0.04}_{-0.04} $& $ \overline{B}_{K^-\eta}\times 10^4 $ & $ 1.34^{+0.04}_{-0.04} $ \\[1ex]
  
 $ \chi^2/\text{d.o.f}$ & $271.5/(182-14)=1.61$ & &\\[1ex]
\hline
\end{tabular}
\caption{The values of parameters from the joint fit. For the values of other relevant parameters, see the text for details. }
\label{tab:fit_results}
\end{table}

\begin{figure}[htbp]
    \centering
    \includegraphics[width=0.6\linewidth]{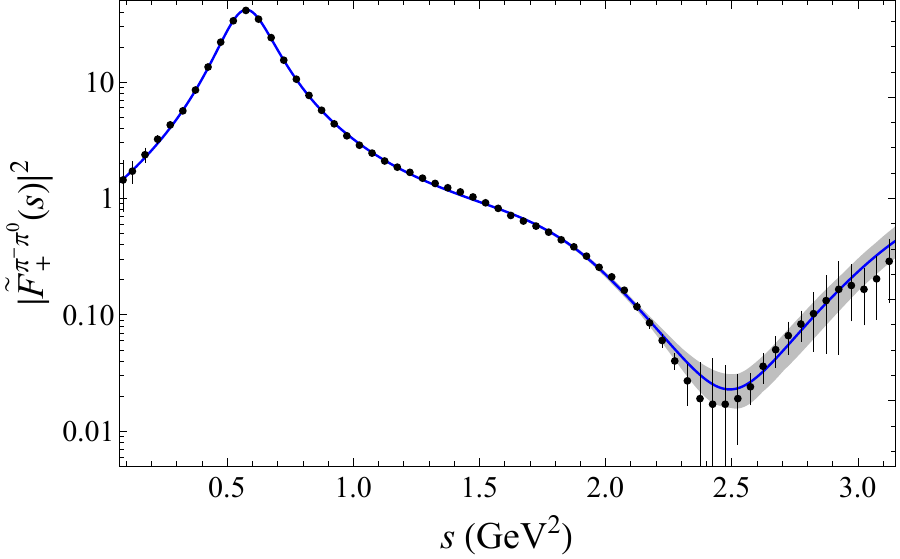}
    \caption{ $\pi\pi$ form factor in $\tau^- \to \pi^-\pi^0 \nu_\tau$. The black filled circles with error bars correspond to data points from Ref.~\cite{Belle:2008xpe}. The blue solid line depicts the modulus squared of the normalized vector form factor $\tilde{F}_{+}^{\pi^-\pi^0}(s)$ as a function of the two-pion energy squared $s$ for the best fit. The gray shaded area represents our estimation of the error band, see the text for detail.}
    \label{pipi}
\end{figure}

In total, we have 14 free parameters and their values resulting from the joint fit, together with error bars, are presented in Table~\ref{tab:fit_results}. 
The fitted values of $M_\rho$, $M_{K^*}$ and $\Gamma_{K^*}$ are consistent with the PDG averages~\cite{ParticleDataGroup:2024cfk}, while the fitted $M_{K^{*'}}$ with $(1.339\pm0.009)$~GeV is smaller than the PDG average with $(1.414\pm 0.015)$~GeV. It is pointed out that similarly smaller values for $M_{K^{*'}}$ are also obtained in the $\tau$ decays in Refs.~\cite{Escribano:2013bca,Escribano:2014joa}. We have further tried other types of fits by fixing the parameters of $M_\rho$, $M_{K^*}$ and $\Gamma_{K^*}$ at their PDG averages, which turn out to marginally affect the results presented in Table~\ref{tab:fit_results}. In contrast, the tentative fit by fixing $M_{K^{*'}}$ at its PDG average leads to obviously increased $\chi^2$. The parameters of $\rho'$ are compatible with the PDG averages within uncertainties. While, the resulting mass of $\rho''$ seems larger than the PDG average and its width is consistent with PDG. Since the $\rho''$ resides in the boundary of the kinematically allowed region of $\tau$ decays, it plays a marginal role. In the following, we will take the values of parameters in Table~\ref{tab:fit_results} to proceed the phenomenological discussions.
We use bootstrap method to estimate the uncertainties of the fitted parameters. Namely, by taking Gaussian sampling of the experimental data, large random pseudo-data sets are generated, which are then used to redo the fits. The large samples of the refitted parameters are exploited to perform all the remaining uncertainty analyses, including the error bars of the parameters, uncertainty bands of the various curves and the predictions of branching ratios that will be addressed later.

\begin{figure}[htbp]
    \centering
    \includegraphics[width=1\linewidth]{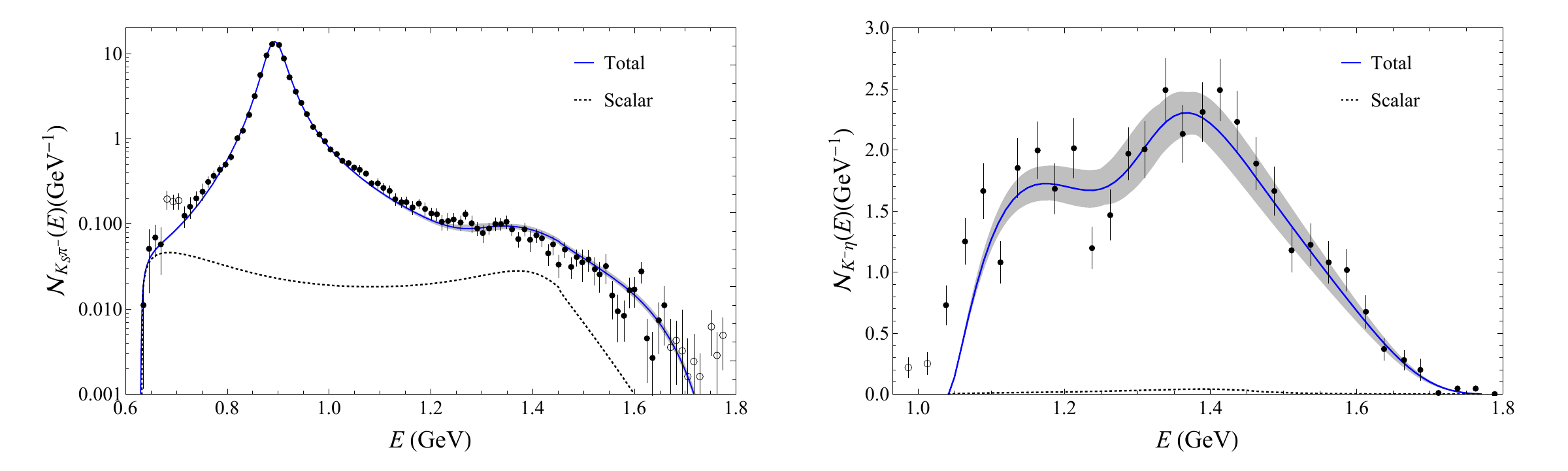}
    \caption{ Normalized event distributions for $\tau^- \to K_S\pi^- \nu_\tau$ (left) and $\tau^- \to K^-\eta\nu_\tau$ (right). The points with error bars are the experimental data from Ref.~\cite{Belle:2007goc} for $K_S\pi^-$ and Ref.~\cite{Belle:2008jjb} for $K^-\eta$. The black filled circles with error bars correspond to the data points included in our fit, while the empty circles denote those excluded from the fit. The blue curves represent the full event distributions from the best fit, whereas black dashed lines denote the contributions from the scalar form factors. The gray shaded areas represent the theoretical error bands.}
    \label{Kpi}
\end{figure}

The curves and the error bands derived from the parameters in Table~\ref{tab:fit_results}, together with the experimental data, are shown in Figs.~\ref{pipi} and \ref{Kpi}, where black solid circles denote experimental data points used in our fit. In Fig. \ref{pipi}, the blue curve represents the modulus squared of normalized vector form factor $\tilde{F}_{+}^{\pi^-\pi^0}(s)= F_{+}^{\pi^-\pi^0}(s)/{\sqrt{2}}$, while in Fig.~\ref{Kpi}, the blue lines depict the total event yield distributions and black dotted lines illustrate the contributions from scalar form factors.

At the end of this section, we present the predictions to the branching ratios by taking our theoretical form factors as inputs for the $\tau^- \to (\pi^-\pi^0,K_S\pi^-,K^-\eta)\nu_\tau$ processes in Table \ref{BRofthreechannels}, along with the separate  contributions from the vector and scalar form factors. Such theoretical predictions to the branching ratios based on form factors are found to be compatible with the fitted normalization factors in Table~\ref{tab:fit_results} for the $K_S\pi^-$ and $K^-\eta$ channels, which reflects the self consistency of our theoretical framework. The theoretical branching ratio for the $\tau^- \to \pi^-\pi^0\nu_\tau$ channel is also nicely consistent with the result of $(25.24\pm 0.39)\%$ from Belle~\cite{Belle:2008xpe}. 

\begin{table}[htbp]
\centering
\renewcommand{\arraystretch}{1.5}  
\begin{tabular}{|>{\centering\arraybackslash}c|>{\centering\arraybackslash}c|>{\centering\arraybackslash}c|>{\centering\arraybackslash}c|}
\hline
Channel & Total & Vector & Scalar\\ [1ex] 
\hline
$\tau^- \to \pi^-\pi^0\nu_\tau$ & $0.254^{+0.002}_{-0.002}$ & $0.254^{+0.002}_{-0.002}$  & -- \\[1ex]
\hline
$\tau^- \to K_S\pi^-\nu_\tau $ & $3.87^{+0.10}_{-0.10} \times10^{-3}$ &  $ 3.78 ^{+0.10}_{-0.10}\times10^{-3}$ &$ 0.88^{+0.105}_{-0.12}\times10^{-4}$\\[1ex]
\hline
$\tau^- \to K^-\eta\nu_\tau $ & $1.24^{+0.13}_{-0.12} \times10^{-4}$ &  $ 1.22 ^{+0.10}_{-0.10} \times10^{-4}$ &$1.48 ^{+2.43}_{-1.42} \times10^{-6}$\\[1ex]
\hline
\end{tabular}
\caption{Theoretical predictions to branching ratios based on our form factors for the $\pi^-\pi^0$, $K_S\pi^-$ and $K^-\eta$ channels in $\tau$ decays.}
\label{BRofthreechannels}
\end{table}

\section{Predictions to the spectra in other channels and the forward-backward asymmetries}\label{sec.prediction}

\begin{figure}[h]
    \centering
    \includegraphics[width=1\linewidth]{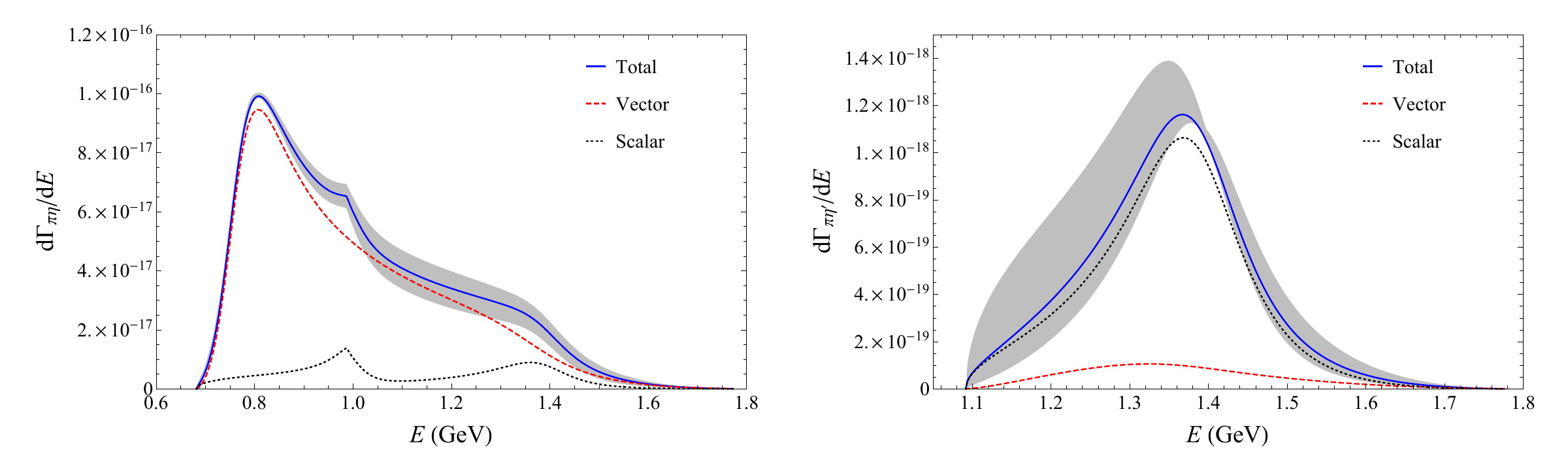}
    \caption{Differential decay widths of $\tau^- \to \pi^-\eta \nu_\tau$ (left) and $\tau^- \to \pi^-\eta'\nu_\tau$ (right) with respect to the energies of the two-meson systems. The blue lines represent total differential decay widths, black dotted lines denote those contributed by the scalar form factors, and red dashed lines indicate those from the vector form factors. The gray shaded areas represent the theoretical error bands.}
    \label{pieta}
\end{figure}

\begin{figure}[h]
    \centering
    \includegraphics[width=0.6\linewidth]{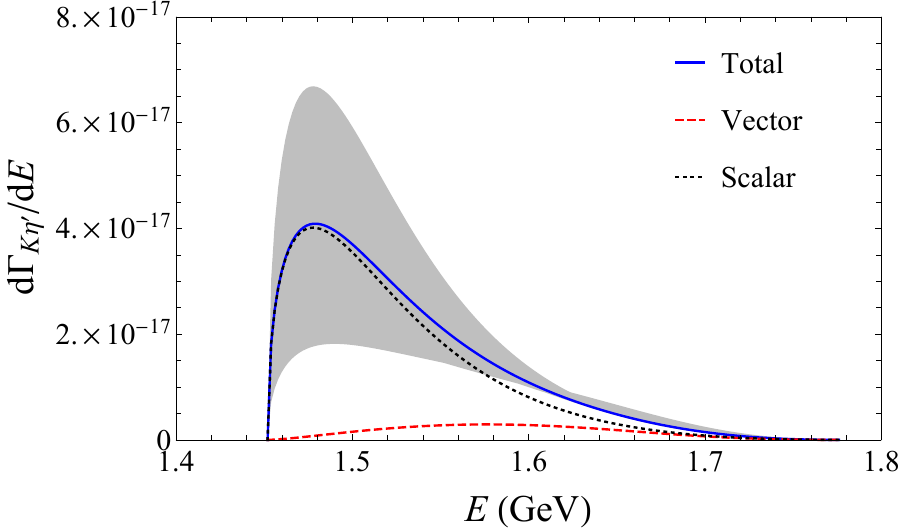}
    \caption{Differential decay width of $\tau^- \to K^-\eta' \nu_\tau$. The meaning of different types of lines is the same with those in Fig.~\ref{pieta}.}
    \label{Ketap}
\end{figure}

\begin{figure}[h]
    \centering
    \includegraphics[width=1\linewidth]{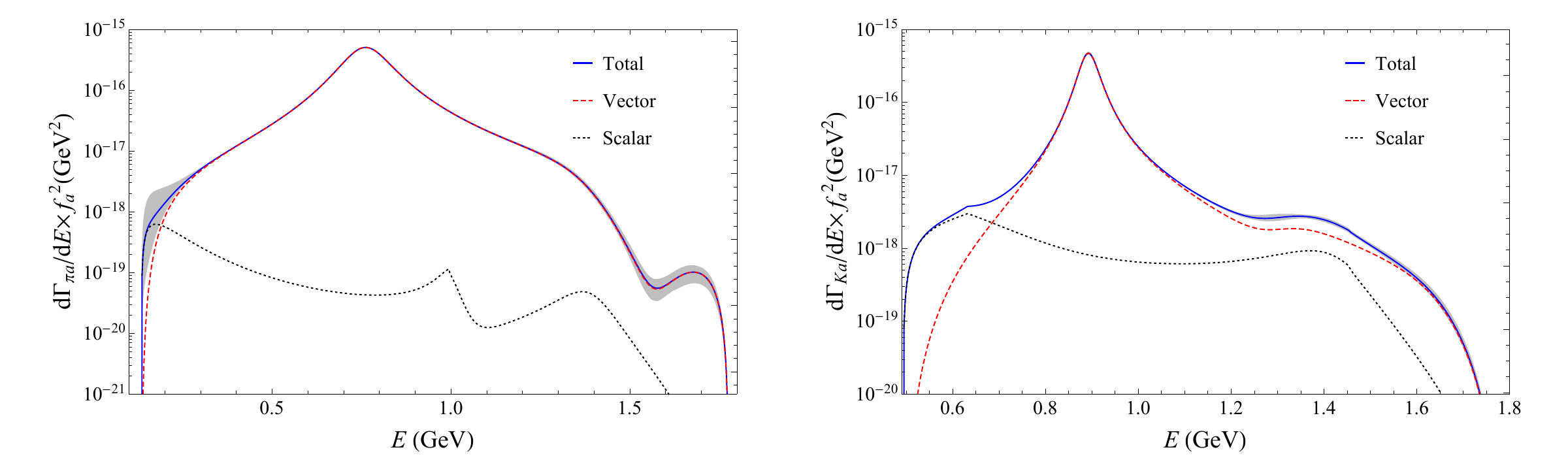}
    \caption{Differential decay widths of $\tau^- \to \pi^- a \nu_\tau$ (left) and $\tau^- \to K^- a \nu_\tau$ (right) by taking $m_a=0$. The meaning of different types of lines is the same with those in Fig.~\ref{pieta}.}
    \label{Ka}
\end{figure}

After fixing all the relevant resonance parameters through the fit elaborated in the previous section, we are ready to present predictions to the differential decay widths for other decay channels, such as $\tau^-\to(\pi^-\eta, \pi^-\eta', K^-\eta', \pi^-a, K^- a)\nu_\tau$, that are not measured yet by experiments. 
The separate contributions from the vector and scalar form factors are also analyzed for different observables. The differential decay widths with respect to the two-boson energies are shown in Fig.~\ref{pieta} for $\pi^-\eta$ and $\pi^-\eta'$,  Fig.~\ref{Ketap} for $K^-\eta'$ and Fig.~\ref{Ka} for $\pi^-a$ and $K^-a$. In the latter figure, we have taken the QCD axion scenario for illustration by setting the bare axion mass $m_{a,0}=0$, and in this case $m_a$ can be set as zero as a perfect approximation in the phenomenological studies~\cite{Gao:2022xqz,Gao:2024vkw}.

According to Figs.~\ref{pieta}-\ref{Ka}, it is obvious that the spectra for the $\pi^-\eta$ and $\pi^-/K^- a$ channels are primarily governed by the vector contributions associated with the $\rho(770)$ and $\rho(770)/K^*(892)$ resonances, respectively. In contrast, the spectra of the $\pi^-\eta'$ and $K^-\eta'$ decay channels are clearly dominated by the scalar form factors. In particular, for the $\pi^-\eta'$ and $K^-\eta'$ spectra, prominent peaks can be easily identified around the energies of $1.4\sim 1.5$~GeV, which correspond to the scalar resonances $a_0(1450)$ and $K_0^*(1430)$, respectively. The $a_0(980)$ manifests as a noticeable kink structure in the $\pi^-\eta$  invariant-mass distribution from the $\tau^-\to\pi^-\eta\nu_\tau$ decay. 
It is interesting to explain the somewhat large uncertainties in the $\tau\to\pi\eta'\nu_\tau$ and $\tau\to K\eta'\nu_\tau$ processes and small error bands for other channels. The reason behind is that the free parameters are fitted to the three types of data sets from the $\tau \to \pi\pi\nu_\tau$, $\tau \to K_S\pi\nu_\tau$ and $\tau \to K\eta\nu_\tau$ processes, which are all dominated by the vector resonances. The scalar resonances do not enter $\tau \to \pi\pi\nu_\tau$ and play marginal roles in the $\tau \to K_S\pi\nu_\tau$ and $\tau \to K\eta\nu_\tau$ decays. 
As a result, the vector resonance parameters are strongly constrained by the data, but the scalar resonance coupling is somewhat loosely constrained in the fit. Consequently, the $\tau\to\pi\eta'\nu_\tau$ and $\tau\to K\eta'\nu_\tau$ processes that are dominated by the contributions from the scalar resonances bear large uncertainties. While, for other channels that are dominated by the vector resonances, the corresponding uncertainties are small.

\FloatBarrier

Additionally, the predicted branching ratios for the five decay channels are tabulated in Table \ref{BRoffivechannels}, where the separate contributions of the vector and scalar form factors to these ratios are given as well. Several recent experimental studies have reported upper limits for some of the decay channels, which are also summarized in Table \ref{BRoffivechannels} for easy comparisons. For the decay mode $\tau^- \to \pi ^-\eta\nu_\tau $,  the BaBar and Belle collaborations presented upper limits of $9.9\times 10^{-5}$~\cite{BaBar:2010bul} at $95\%$ confident level (C.L.) and $7.3 \times 10^{-5}$~\cite{Hayasaka:2009zz} at $90\%$ C.L., respectively. Our theoretical model predicts the branching ratio of this channel to be $1.63^{+0.14}_{-0.14} \times10^{-5}$, which is in good agreement with the existing experimental upper limits. For the $\tau^- \to \pi^-\eta'\nu_\tau $ decay mode, the BaBar collaboration has also made some progress, updating the upper limit to be $4.0 \times 10^{-6}$~\cite{BaBar:2012zfq} at $90\% $ C.L., which outperforms its previous result of $7.2 \times 10^{-6}$~\cite{BaBar:2008wlm} at $90\% $ C.L. Our predicted branching ratio for this channel is $1.17^{+0.36}_{-0.07} \times10^{-7}$, which is much lower than the current experimental limit. As for the $\tau^- \to K^-\eta'\nu_\tau $ decay mode, the BaBar collaboration~\cite{BaBar:2012zfq} has provided a measurement, determining an upper branching ratio limit as $2.4 \times 10^{-6}$ at $90\%$ C.L. Our theoretical prediction for this channel is about $2.00^{+1.01}_{-0.70} \times10^{-6}$, which is around the edge of the experimental limit. 
Regarding the axion production rates, we verify that the hadronic resonance contributions give significant enhancements, compared to the case by only including the LO chiral results. Taking the situation of $m_a=0$ for illustration, the enhanced ratio of the $\tau\to\pi a\nu_\tau$ decay widths by using the complete hadronic amplitude and the LO chiral amplitude is found to be 7.5, while such ratio in the $\tau\to K a\nu_\tau$ channel is enlarged to be 19.2.

\begin{table}[htbp]
\centering
\renewcommand{\arraystretch}{1.5}  
\begin{tabular}{|>{\centering\arraybackslash}c|>{\centering\arraybackslash}c|>{\centering\arraybackslash}c|>{\centering\arraybackslash}c|c|}
\hline
 Channel &Total &Vector &Scalar & Exp Limits\\  [1ex] 
\hline
\raisebox{-0.7ex}{\parbox{0.15\textwidth}{\centering $\tau^- \to \pi^-\eta\nu_\tau$ \\[0.5ex]
                             ($\times 10^5$) }} 
                              
& $1.63^{+0.14}_{-0.14}$ &  $ 1.43^{+0.18}_{-0.21} $ &$0.20 ^{+0.07}_{-0.04}$ 
& 
\raisebox{-0.7ex}{\parbox{0.2\textwidth}{\centering $< 9.9$ (BaBar)~\cite{BaBar:2010bul} \\[0.5ex]
                              $<7.3 $ (Belle)~\cite{Hayasaka:2009zz}}} 
\\[1ex] 
\hline
\raisebox{-0.7ex}{\parbox{0.15\textwidth}{\centering $\tau^- \to \pi^-\eta'\nu_\tau$  \\[0.5ex]
                             ($\times 10^7$) }} 
& $1.17^{+0.36}_{-0.07} $ &  $ 0.14^{+0.09}_{-0.08}$ &$1.03 ^{+0.44}_{-0.16}$
&  $< 40 $ (BaBar)~\cite{BaBar:2012zfq} \\[1ex]
\hline
\raisebox{-0.7ex}{\parbox{0.15\textwidth}{\centering$\tau^- \to K^-\eta'\nu_\tau $ \\[0.5ex]
                             ($\times 10^6$) }} 
& $2.00^{+1.01}_{-0.70}$ &  $ 0.21 ^{+0.34}_{-0.14} $ &$1.79 ^{+1.15}_{-1.04}$ 
& $< 2.4 $ (BaBar)~\cite{BaBar:2012zfq} 
\\[1ex]
\hline
\raisebox{-0.7ex}{\parbox{0.15\textwidth}{\centering $\tau^- \to \pi^- a\nu_\tau $ \\[0.5ex]
                             ($\times 10^5 $ GeV$^{2}$) }} 
& $4.44^{+0.04}_{-0.03}/f_a^2$ &  $ 4.44 ^{+0.04}_{-0.03} / f_a^2$ &$0.006^{+0.011}_{-0.004}/f_a^2$
& -- \\[1ex]
\hline
\raisebox{-0.7ex}{\parbox{0.15\textwidth}{\centering $\tau^- \to K^- a\nu_\tau $ \\[0.5ex]
                             ($\times 10^5$ GeV$^{2}$) }}  & $1.48^{+0.04}_{-0.04}/f_a^2$ &  $ 1.43 ^{+0.04}_{-0.04}/f_a^2$ &$0.05 ^{+0.01}_{-0.01} /f_a^2$
& --\\[1ex]
\hline
\end{tabular}
\caption{Theoretical predictions to branching ratios for the $\pi^-\eta/\eta' $, $K^-\eta'$, $\pi^-a$ and $K^- a $ channels in $\tau$ decays. Contributions of corresponding vector and scalar form factors are separately given. The experimental upper limits are also provided if available.}
\label{BRoffivechannels}
\end{table}

\begin{figure}[h]
    \centering
    \includegraphics[width=1\linewidth]{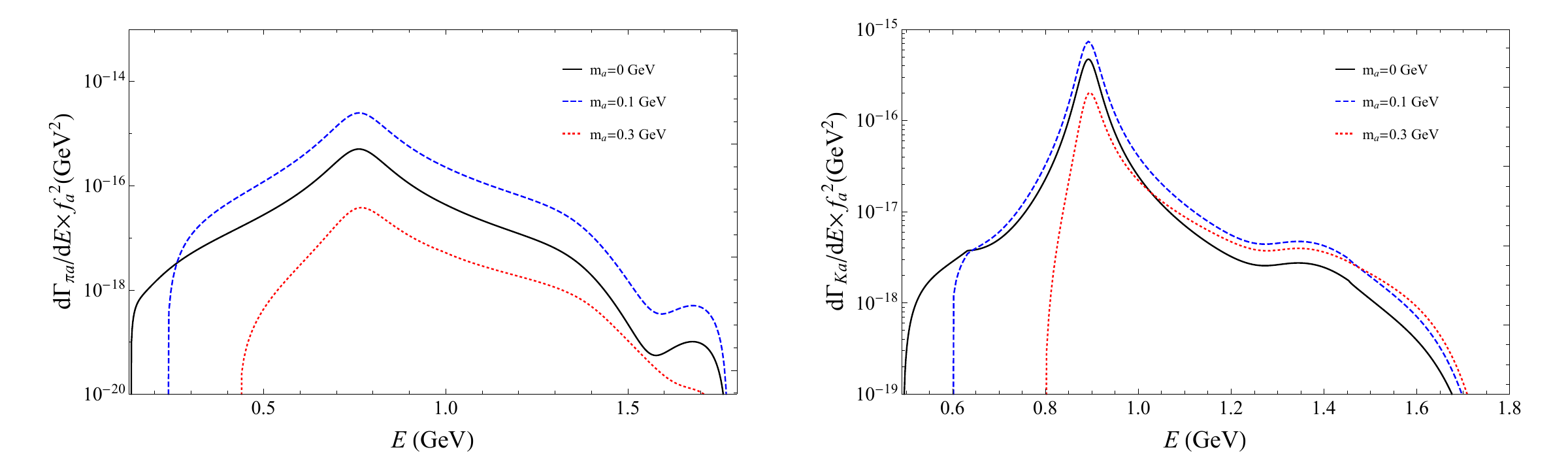}
    \caption{Differential decay widths of $\tau^- \to \pi^- a \nu_\tau$ (left) and $\tau^- \to K^- a \nu_\tau$ (right) for ALP masses at $m_a=0.1$ and 0.3~GeV. The QCD axion case with $m_a=0$ is shown for comparison. }
    \label{piama}
\end{figure}

\begin{figure}[h]
    \centering
    \includegraphics[width=1\linewidth]{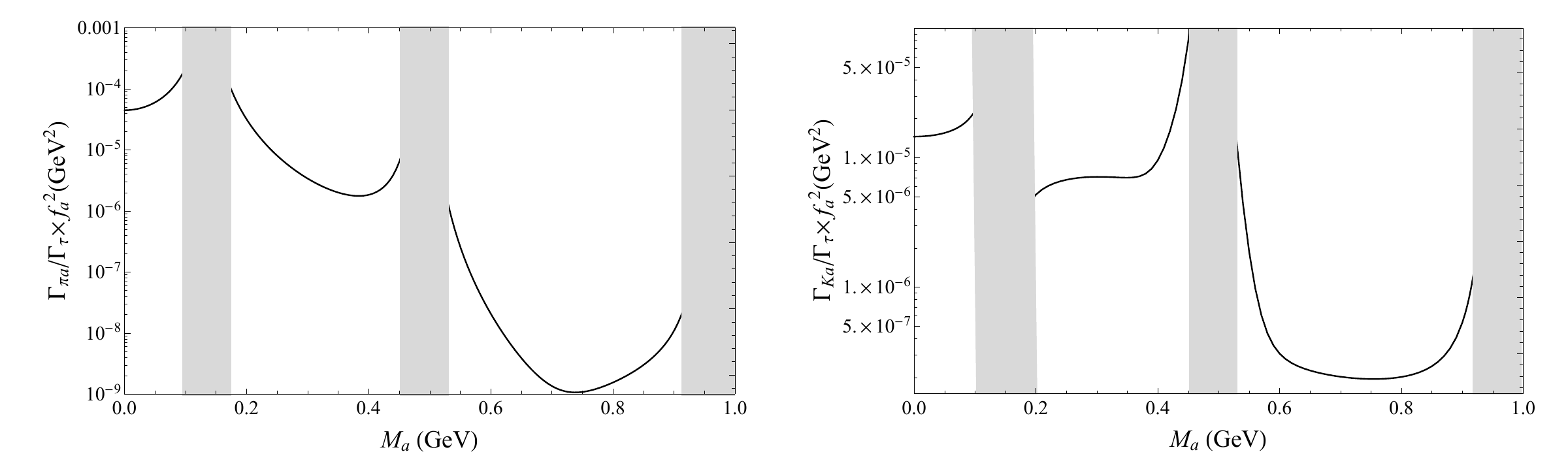}
    \caption{Predicted branching ratios of $\tau^- \to \pi^- a \nu_\tau$ (left) and $\tau^- \to K^- a \nu_\tau$ (right) as a function of the ALP mass $m_a$. The gray bands mask regions near $m_a \sim m_{\overline{\pi}},\, m_{\overline{\eta}},\,m_{{\overline{\eta}'}}$, merely reflecting that the perturbative treatment of axion-meson mixing around these regions is deficient. }
    \label{Gammapiadifferentma}
\end{figure}

Apart from the QCD axion case, we further explore the phenomenological consequences by introducing the bare mass term $m_{a,0}$ for the axion, which is usually deemed as the axion-like particle (ALP) scenario. In this situation, the ALP mass $m_a$ is predominated by the bare mass $m_{a_,0}$. 
We present the differential decay widths of $\tau^- \to \pi^-/K^-a \nu_\tau$ channels by taking $m_a=0.1$ and 0.3~GeV in Fig. \ref{piama}, where the curves corresponding to $m_a=0$ are also included for comparison. 
By continuously varying the ALP masses from 0 to 1~GeV, we give predictions to the branching ratios of the $\tau^- \to \pi^-/K^-a \nu_\tau$ processes as a function of $m_a$ in Fig.~\ref{Gammapiadifferentma}. 
The singularities in the latter figure emerge at specific mass thresholds, i.e., when $m_a \sim m_{\overline{\pi}},\, m_{\overline{\eta}},\,m_{{\overline{\eta}'}}$, due to the fact that the axion-meson mixing matrix elements $v_{4i/i4}$ and $y_{i4}$ (with $i=1,2,3$) are proportional to $\frac{1}{m_a^2 - m_i^2}$,  with $m_i= m_{\overline{\pi}}, m_{\overline{\eta}}, m_{\overline{\eta}'}$, according to Refs.~\cite{Gao:2022xqz,Gao:2024vkw}. It is pointed out that we have taken the same LO isospin-limit masses in the denominators of both $v_{4i/i4}$ and $y_{i4}$ here, while the isospin breaking corrected masses are used for $y_{i4}$ in Ref.~\cite{Gao:2024vkw}. The seemingly magnificent enhancements around $ m_a = m_{\overline{\pi}},\, m_{\overline{\eta}},\,m_{{\overline{\eta}'}}$ in Fig.~\ref{Gammapiadifferentma} are expected to be artificial, merely reflecting the deficiency of the perturbative treatment of the axion-meson mixing around these regions, and therefore are shaded in the plots. As a result, only the predictions distant from the regions of $m_a = m_{\overline{\pi}},\, m_{\overline{\eta}},\,m_{{\overline{\eta}'}}$ are considered to be meaningful. 

It should be noted that our study presents a complementary calculation of the ALP production rates in the tau decays to the work of Ref.~\cite{Ema:2025bww}, since the latter reference only incorporates the ALP-lepton coupling, while ours assumes that the predominant mechanism in the $\tau\to\pi(K) a \nu_\tau$ processes is the model-independent $aG\tilde{G}$ interaction. According to the branching ratios of the ALP production from the $\tau$ decays illustrated in Fig.~\ref{Gammapiadifferentma}, although it could be difficult to observe such $\tau$ decay processes with tiny absolute branching ratios in the current experiments, our study reveals an alternative interesting feature that the ratios of $\frac{\Gamma_{\tau\to K a\nu_\tau}/V_{us}^2}{\Gamma_{\tau\to \pi a\nu_\tau}/V_{ud}^2}\simeq (6.3_{(m_a=0)}, \, 41_{(m_a=0.3~\rm{GeV})}, \, 2513_{(m_a=0.8~\rm{GeV})} )$ are much larger than those of $\frac{\Gamma_{\tau\to K \nu_\tau}/V_{us}^2}{\Gamma_{\tau\to \pi \nu_\tau}/V_{ud}^2}\simeq 1.2$ and $\frac{\Gamma_{\tau\to K \pi\nu_\tau}/V_{us}^2}{\Gamma_{\tau\to \pi \pi\nu_\tau}/V_{ud}^2}\simeq 0.3$. This significant enhancement in the $aK$ system, compared to the $\pi a$ one, is in accord with the recent finding in Ref.~\cite{Wang:2025wlu} that explores the $a K\to \pi K$ and $a\pi \to \pi\pi$ interactions. We expect that such enhanced ratio of $\frac{\Gamma_{\tau\to K a\nu_\tau}/V_{us}^2}{\Gamma_{\tau\to \pi a\nu_\tau}/V_{ud}^2}$ will not persist in the axion-lepton dominant scenario in Ref.~\cite{Ema:2025bww}. Therefore this ratio could provide a useful quantity to distinguish different theoretical axion models. Interestingly, the authors of Ref.~\cite{Ema:2025bww} show that their branching ratio of $\tau\to\pi a\nu_\tau$ is around $10^{-8}|g_{a\tau\tau}|^2$ by setting $f_a=100$~GeV, which is roughly similar to ours when taking the axion-tau coupling $g_{a\tau\tau}$ as $O(1)$. 
As illustrated in Ref.~\cite{Ema:2025bww}, in order to set realistic constraints on the ALP parameters, one would also need the ALP decay information as inputs to match the experimental conditions, which is beyond the scope of the present study. We believe that such an analysis definitely deserves another independent work in future. It is reiterated that one of the key novelties of this work is to systematically include the hadronic resonance contributions to the ALP production in tau decays with the model-independent interaction of the $aG\tilde{G}$ type.

\FloatBarrier

\begin{figure}[h]
    \centering
    \includegraphics[width=1\linewidth]{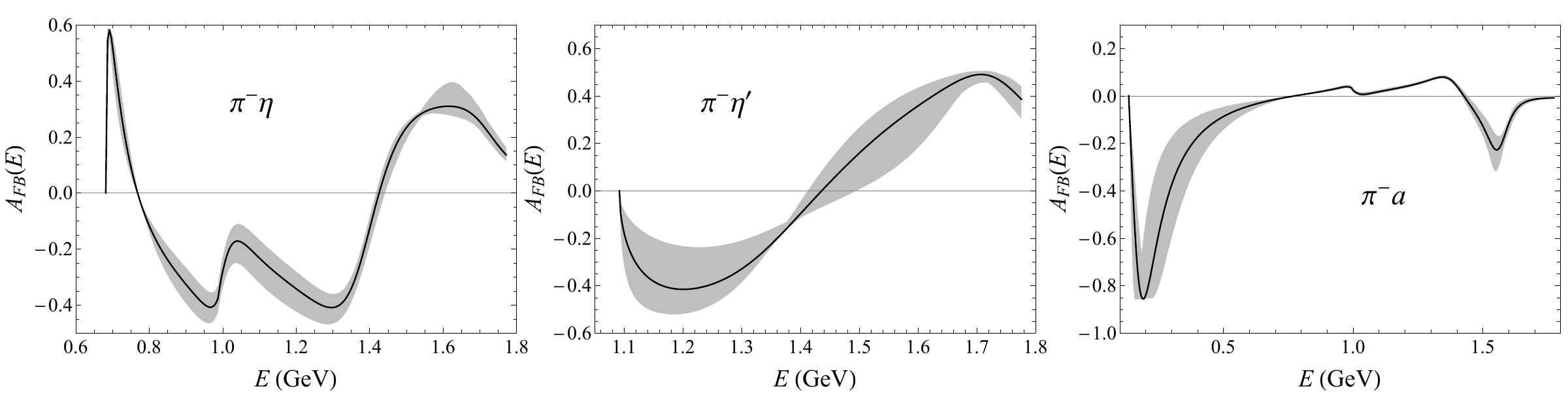}
    \caption{From left to right, the plots show $A_{FB}$ for the $\tau^- \to \pi^-\eta\nu_\tau$, $\tau^- \to \pi^-\eta^\prime \nu_\tau$, and $\tau^- \to \pi^-a\nu_\tau$ channels as a function of the invariant mass of final-state pseudoscalar bosons. The gray shaded areas represent our estimation of theoretical error bands.}
    \label{piPABF}
\end{figure}

\begin{figure}[h]
    \centering
    \includegraphics[width=0.9\linewidth]{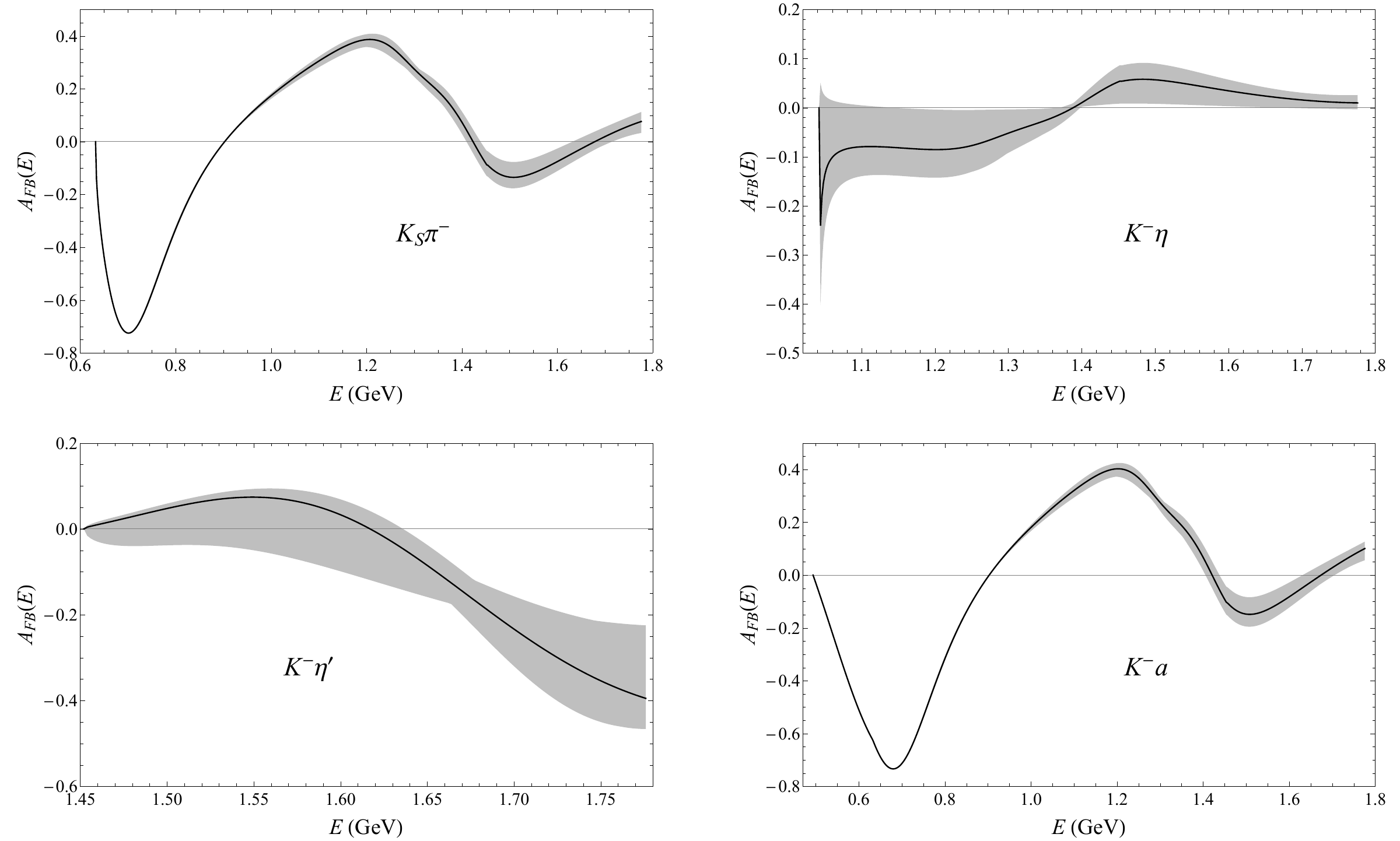}
    \caption{From top-left to bottom-right: $A_{FB}$ for $\tau^- \to K_S\pi^-\nu_\tau$, $\tau^- \to K^-\eta\nu_\tau$, $\tau^- \to K^-\eta^\prime \nu_\tau$ and $\tau^- \to K^-a\nu_\tau$ as a function of the invariant mass of final-state pseudoscalar bosons. The gray shaded areas represent our estimation of theoretical error bands.}
    \label{KpiABF}
\end{figure}

Last but not least, we present predictions to the forward-backward asymmetries arising from the two-pseudoscalar boson decays of the $\tau$ lepton, viz., $A_{FB}$ defined in Eq.~\eqref{eq.afb}. The FB asymmetry $A_{FB}$ provides a valuable quantity to probe the interference term between the vector and scalar form factors, which is absent in the differential decay width with respect to the two-boson energy given in Eq.~\eqref{dGammapi}. 
For the decay channel $\tau^- \to \pi^-\pi^0\nu_\tau$, as discussed earlier, the contribution of the scalar form factor to this channel is much suppressed, compared to the vector form factor, and therefore it is neglected in our study, indicating that $A_{FB}$ in this channel is zero. Regarding the FB asymmetries for the other channels, the resulting curves are given in Fig.~\ref{piPABF} for the Cabibbo allowed channels, including $\tau^- \to (\eta,\eta',a)\pi^-\nu_\tau$, and Fig.~\ref{KpiABF} for the Cabibbo suppressed channels, such as $\tau^- \to (K_S\pi^-,K^-\eta,K^-\eta',K^-a)\nu_\tau$.

\FloatBarrier

\section{Summary and conclusions}\label{sec.sum}

In this work, we exploit resonance chiral theory augmented with model-independent axion interaction operator $aG\tilde{G}/f_a$ to calculate the form factors for the two-meson and axion-meson production from the semileptonic $\tau$ decays, by utilizing the $\pi$-$\eta$-$\eta'$-$a$ four-particle mixing matrix elements in Ref.~\cite{Gao:2024vkw}. Our calculations simultaneously include the following channels for both Cabibbo-allowed and Cabibbo-suppressed cases: $\tau^- \to \pi^- \pi^0\nu_\tau,\tau^- \to \pi^- \eta/\eta'\nu_\tau, \tau^- \to K_S \pi^-\nu_\tau,\tau^- \to K^- \eta/\eta'\nu_\tau$ and $ \tau^- \to \pi^-/K^- a\nu_\tau$. 

We have successfully performed the joint fit by reasonably reproducing the experimental data from the Belle collaboration on the two-meson invariant-mass spectra in the $\tau^- \to \pi^- \pi^0 \nu_\tau$~\cite{Belle:2008xpe}, $\tau^- \to K_S \pi^-\nu_\tau$~\cite{Belle:2007goc} and $\tau^- \to K^- \eta\nu_\tau$~\cite{Belle:2008jjb}, where the relevant unknown resonance parameters are determined. 
We then utilize those fitted resonance parameters to make predictions to the two-boson spectra and branching ratios for other five decay channels, including $\tau^- \to \pi^- \eta/\eta'\nu_\tau,\tau^- \to K^- \eta'\nu_\tau$ and $ \tau^- \to \pi^-/K^- a\nu_\tau$. 
Apart from the QCD axion scenario, we extend our analysis to the axion-like particle case by introducing a nonzero axion mass $m_{a}$. We compute differential decay widths for $ \tau^- \to \pi^-/K^- a\nu_\tau$ channels at $m_{a}=0$, 0.1~GeV and 0.3~GeV for illustrations, and calculate these branching ratios as a function of $m_{a}$ across a continuous range from 0 to 1~GeV. Our study reveals that to properly include the hadronic resonance contributions in the axion/axion-like particle productions from the tau decays are critical, since the production rates can be enlarged by around one order of magnitude, compared to the case by only including the leading-order chiral amplitudes.

Our predicted branching ratios to $\tau^- \to \pi^- \eta/\eta'\nu_\tau$ and $\tau^- \to K^-\eta'\nu_\tau$ are in agreement with the current experimental upper limits. 
Furthermore, we present predictions to forward-backward asymmetries in the two-pseudoscalar boson decays of the $\tau$ lepton, which probe vector-scalar form factor interference that is absent in the differential decay widths with respect to the two-boson energy. 
Our predictions to the branching ratios, invariant-mass distributions and the forward-backward asymmetries can provide useful guidelines to the future experimental measurements, such as the ongoing Belle II and future facilities like the Super Tau-Charm Facility and the Circular Electron Positron Collider.

\section*{Acknowledgements}

This work is supported in part by National Natural Science Foundation of China (NSFC) under Grants No.~12475078, No.~12150013, No.~11975090, and also by the Science Foundation of Hebei Normal University with Contract No.~L2023B09. 

\bibliography{tau-pa}
\bibliographystyle{apsrev4-2}

\end{document}